\documentclass[%
 reprint,
 amsmath,amssymb,
 aps,
 nofootinbib
]{revtex4-2}

\usepackage{xspace}
\usepackage{graphicx}
\usepackage{dcolumn}
\usepackage{bm}
\usepackage{xcolor}
\usepackage[normalem]{ulem}
\usepackage[colorlinks,bookmarks]{hyperref}
\usepackage{orcidlink}
\definecolor{linkblue}{rgb}{0,0,0.8}
\definecolor{linkgreen}{rgb}{0,0.5,0}

\hypersetup{pdfpagemode=UseNone, pdfstartview=FitH, linkcolor=linkblue, 
	citecolor=linkgreen, urlcolor=linkblue}

\newcommand{\tcm}{21$\,$cm\xspace}  
\newcommand{\thetaEFT}{\bm{\theta}_{\rm EFT}}
\newcommand{\thetaHOD}{\bm{\theta}_{\rm HOD}}
\newcommand\numberthis{\addtocounter{equation}{1}\tag{\theequation}}

\newcommand{\HI}{H$\textsc{i}$}

\begin{document}


\title{Simulation-Based Priors for HI Bias from Halo Occupation Physics}

\author{Debanjan Sarkar\,\orcidlink{0000-0001-5763-2541}}
\email{debanjan.sarkar@mcgill.ca}
\affiliation{\,Department of Physics and Trottier Space Institute, McGill University, QC H3A 2T8, Canada}
\affiliation{\,Ciela—Montreal Institute for Astrophysical Data Analysis and Machine Learning, QC H2V0B3, Canada}

\author{Simon Foreman\,\orcidlink{0000-0002-0190-2271}}
\affiliation{\,Department of Physics, Arizona State University, Tempe, AZ 85287, USA}

\date{\today}

\begin{abstract}
Full-shape analyses of 21\,cm intensity maps with the effective field 
theory of large-scale structure will require priors on \HI\ bias 
parameters, and the standard choice of broad uninformative priors
can lead to cosmological constraints that are unnecessarily conservative.
We present a simulation-based 
framework that replaces these broad priors with informative priors based on learning the conditional distribution 
$p(\bm{\theta}_{\rm EFT}\mid\bm{\theta}_{\rm HOD})$ between 
effective-field-theory-based bias parameters and the parameters of a model for \HI\ clustering in the nonlinear regime. 
Specifically, we train a conditional normalizing flow on field-level measurements of the lowest-order local bias parameters $(b_1,b_2,b_3)$ and the tidal bias $b_{\mathcal{G}_2}$ by applying a simple \HI\ halo occupation distribution (HOD) to the Hidden Valley simulations.
We find that the resulting HOD-to-bias mapping is highly structured,
displaying a strong dependence on the power of halo mass in the HOD model.
Propagating CHORD-like telescope sensitivity forecasts for the \tcm power spectrum on nonlinear scales through this 
mapping produces non-Gaussian, correlated priors on the bias 
parameters that are substantially tighter than conventional flat 
priors across $z=1$--$3$, with the improvement most dramatic at 
high redshift. 
By repeating our analysis using halo catalogs from the IllustrisTNG simulations,
we find non-negligible differences from the Hidden Valley results, indicating that
future applications of simulation-based \HI\ priors will need to carefully account for
the dependence of these priors on the simulations used to construct them.
Our framework provides an initial step toward informative EFT priors for current and forthcoming \HI\ intensity mapping surveys,
including CHIME, CHORD, and MeerKLASS.
\end{abstract}

\maketitle

\section{Introduction}\label{sec:intro}

Neutral-hydrogen (\HI) intensity mapping with the redshifted \tcm line offers a route to precision cosmology over large volumes in the post-reionization universe. Rather than resolving individual galaxies, it measures aggregate line emission from unresolved systems, probing three-dimensional large-scale structure without a detection threshold. This makes it a natural complement to galaxy redshift surveys and well suited to cross-correlation analyses. Indeed, several cross-correlations have been successfully detected, including with galaxy surveys~\cite{Pen:2008fw,Chang:2010jp,Masui:2012zc,Anderson:2017ert,Li:2020pre,Tramonte:2020csa,eBOSS:2021ebm,CHIME:2022kvg,Cunnington:2022uzo,Carucci:2024qpm,Chen:2025bcx} and the Lyman-$\alpha$ forest~\cite{CHIME:2023til}, while non-Gaussian cross-correlations with gravitational lensing of the cosmic microwave background are being explored with data~\cite{CHIME:2026rbk}. Furthermore, \tcm auto spectrum analyses have been carried out using data from MeerKAT~\cite{Paul:2023yrr,Townsend:2026wec} and the Canadian Hydrogen Intensity Mapping Experiment (CHIME)~\cite{CHIME:2025cee,CHIME:2026htj}. For broader introductions to 21\,cm and line-intensity mapping, see, e.g., Refs.~\cite{Liu:2019awk, Kovetz:2017agg, Chang:2026ake}.

Turning intensity maps into cosmological constraints requires models that remain accurate beyond strictly linear scales. At post-reionization redshifts, the \HI\ signal depends on mildly to very nonlinear structure formation and on the astrophysics governing how neutral gas occupies halos \cite{Sarkar:2016lvb, Sarkar:2019ojl}. 
The effective field theory of large-scale structure (EFT of LSS~\cite{Baumann:2010tm}) handles the quasi-linear regime by organizing biased-tracer predictions in a basis of long-wavelength operators, with short-scale physics absorbed into a finite set of bias parameters and counterterms. In this language, the large-scale \HI\ density field is characterized by a set of bias parameters (including higher-order and tidal biases) together with additional nuisance parameters in real and redshift space; see, e.g., Ref.~\cite{Desjacques:2016bnm} for a review.

A natural consequence of this approach is that the values of the EFT parameters that enter full-shape power spectrum analyses are very uncertain \textit{a priori}. In previous applications of EFT predictions to large-scale structure, this uncertainty is typically handled by adopting broad priors on EFT parameters and marginalizing. This is conservative, but costly, as loose priors can significantly soften cosmological constraints.

In the galaxy-clustering literature, a sharper approach has been developing: build \emph{simulation-based priors} (SBPs) for EFT parameters by generating controlled mock catalogs spanning the parameter space of halo occupation models, measuring the corresponding EFT parameters, and learning their joint distribution. That distribution then serves as an informative prior in survey likelihood analyses. 
This approach has been applied to full-shape analyses of BOSS~\cite{Ivanov:2024hgq,Ivanov:2024xgb} and DESI~\cite{DESI:2025wzd,Chen:2025jnr,Chudaykin:2026nls}, and explored in various ways using simulations~\cite{Zhang:2024thl,Akitsu:2024lyt,Ivanov:2024dgv,Shiferaw:2024ehr,Ivanov:2025qie,Zennaro:2021pbe}.

For \HI, the key enabling ingredient is field-level forward modeling that can be validated directly against simulations. In Refs.~\cite{Obuljen:2022cjo,Foreman:2024kzw}, a perturbative forward model with a bias expansion was shown to reproduce post-reionization \HI\ overdensity maps from the IllustrisTNG simulations~\cite{TNGa,TNGb,TNGc,TNGd,TNGe} (in real and redshift space), and to yield precise bias measurements by making use of sample variance cancellation. This provides a practical route to measuring \HI\ bias parameters across large ensembles of modest-volume \HI\ realizations.

This paper presents a first attempt at an \HI-focused SBP framework to answer a concrete question: how do variations in small-scale \HI\ halo-occupation physics, as approximated by a halo occupation distribution (HOD) model, map into the large-scale EFT bias parameters that enter perturbative full-shape analyses? 
We do so by generating an ensemble of \HI\ maps, based on using a compact \HI--HOD parameterization to assign \HI\ masses to halos from the public Hidden Valley $N$-body simulations~\cite{Modi:2019ewx}.
For each map, we infer several bias parameters---the local biases $(b_1,b_2,b_3)$ and the tidal bias $b_{\mathcal G_2}$---for each realization using a field-level forward-modeling pipeline, and then model the resulting conditional distribution with normalizing flows. This gives a sampleable map from HOD parameters $\thetaHOD$ to EFT parameters $\thetaEFT$ that can seed SBPs for full-shape analyses, enabling an exploration of the relationship 
the two sets of parameters.

We find that this mapping is highly structured: the EFT bias parameters do not vary independently across the simulated \HI\ ensemble, but instead occupy a curved, correlated, and non-Gaussian region of $\thetaEFT$ space. This structure is precisely the information that is lost when broad, independent priors are assigned to each bias parameter.

The learned conditional density can be used in two complementary ways. One option is to construct a ``pure'' simulation-based prior by sampling over an assumed distribution of $\thetaHOD$ and marginalizing the learned density over the HOD parameters; this is the closest analogue of existing SBP applications in galaxy surveys. A second option is to use external or nonlinear-scale measurements to constrain $\thetaHOD$, and then propagate those constraints through the flow to obtain an induced prior on $\thetaEFT$ for quasi-linear EFT analyses. As a proof of concept of this second use case, we 
forecast 21\,cm power spectrum constraints from a survey inspired by the specifications of the Canadian Hydrogen Observatory and Radio-transient Detector (CHORD~\cite{Vanderlinde:2019tjt}), and propagate them through the trained flow. We find 
informative, correlated, and in some cases visibly non-Gaussian priors on $(b_1,b_2,b_{\mathcal G_2},b_3)$. We also compare the learned mappings from Hidden Valley and IllustrisTNG300, finding that the qualitative structure is similar while quantitative offsets remain, especially in the high-bias and tidal sectors.

The paper is organized as follows. Section~\ref{sec:theory} presents the field-level \HI\ modeling and bias-inference pipeline, and Section~\ref{sec:hv-results} presents the results of applying this pipeline to ensembles of \HI\ maps generated from the Hidden Valley simulations.
Section~\ref{sec:flows} describes the conditional normalizing-flow model and demonstrates how it captures the Hidden Valley HOD-to-bias mapping. Section~\ref{sec:tng} discusses the dependence of the results on the underlying simulation, by repeating our procedure using the \texttt{TNG300} run of the IllustrisTNG simulations. 
Section~\ref{sec:flow_usage} contains forecasts for a CHORD-like 21\,cm survey that use our normalizing flows to derive bias-parameter priors from measurements of the 21\,cm power spectrum on nonlinear scales.
We summarize and outline next steps in Sec.~\ref{sec:summary}.
A set of appendices contains additional details and validation checks.

\section{Field-level HI modelling and bias inference}\label{sec:theory}

This section describes the field-level framework used to (i) construct \HI\ overdensity maps from halo catalogs and (ii) infer large-scale \HI\ bias parameters from those maps using a perturbative forward model. The methodology follows the shifted-operator and transfer-function approach first developed for dark matter halos in
Refs.~\cite{Schmittfull:2018yuk,Schmittfull:2020trd}
and applied to \HI\ in
Refs.~\cite{Obuljen:2022cjo,Foreman:2024kzw}.

\subsection{Halo catalogues and construction of HI fields}\label{sec:theory_halo_hi}

The \emph{Hidden Valley} (HV) simulations~\cite{Modi:2019ewx}\footnote{\url{https://cyril.astro.berkeley.edu/HiddenValley/}} are a high-resolution dark-matter-only $N$-body suite designed for post-reionization \HI\ intensity-mapping applications. In the flagship configuration, \texttt{HV10240} evolves $10240^3$ particles in a periodic cube of side length $L=1024\,h^{-1}{\rm Mpc}$, with particle mass $m_{\rm p}=8.58\times 10^{7}\,h^{-1}M_\odot$ and mean inter-particle spacing 
$100\,h^{-1}{\rm kpc}$. The simulations are run with \texttt{FastPM}~\cite{Feng:2016yqz} using a $20480^3$ particle-mesh for gravity and 35 time steps from $z=99$ to $z=0.5$; initial conditions are generated with 2LPT at $z=99$ using a linear power spectrum from \texttt{CAMB}~\cite{Lewis:1999bs}. The fiducial cosmology is close to the final \emph{Planck} values: 
$\Omega_{\rm m}=0.309167$, $\Omega_{\rm b}=0.04903$, $h=0.677$, $n_s=0.96824$, $\sigma_8=0.8222$. 
For our analysis, we use snapshots at seven redshifts spanning $z=1$ to $z=4$ in steps of 
$\Delta z = 0.5$, i.e.\ $z\in\{1.0,\,1.5,\,2.0,\,2.5,\,3.0,\,3.5,\,4.0\}$, covering the primary observing windows of CHIME~\cite{CHIME:2022dwe}, CHORD~\cite{Vanderlinde:2019tjt} and SKA-Mid~\cite{SKA:2018ckk}.

To investigate the dependence of our results on the underlying simulation, we also use the public halo and subhalo catalogues from the largest run from the IllustrisTNG simulations~\cite{TNGa,TNGb,TNGc,TNGd,TNGe}, \texttt{TNG300} ($L=205\,h^{-1}{\rm Mpc}$; $2\times2500^3$ resolution elements). 
The dark-matter particle mass is $m_{\rm DM}\simeq 5.9\times10^7\,M_\odot$ and the target baryonic (gas cell) mass is $m_{\rm b}\simeq 1.1\times10^7\,M_\odot$. \texttt{TNG300} adopts the Planck 2015 cosmology~\citep{Planck:2015fie}: $\Omega_{\rm M}=0.3089$, $\Omega_{\rm b}=0.0486$, $h=0.6774$, $n_s=0.9667$, $\sigma_8=0.8159$, which is close to but distinct from the HV fiducial cosmology; we account for this difference where relevant in our comparisons. Dark-matter halos are identified with a friends-of-friends (FoF) algorithm (linking length $b=0.2$) applied to dark-matter particles, with gas, stars, and black holes attached to the same FoF group as their nearest dark-matter particle. 
Gravitationally bound substructures are then identified with \texttt{Subfind}, which returns both central and satellite subhalos within each FoF halo. We analyse the same seven redshift snapshots used for the HV suite, $z\in\{1.0,\,1.5,\,2.0,\,2.5,\,3.0,\,3.5,\,4.0\}$, all of which are available in the public \texttt{TNG300} data release\footnote{\url{https://www.tng-project.org/data/downloads/TNG300-1/}}. In this work we 
use only halo positions and masses from these catalogues, and present our HV--TNG comparisons in Sec.~\ref{sec:tng}.

Given a halo of mass $M$, we assign a mean \HI\ mass using the \HI--HOD parameterization~\cite{Villaescusa-Navarro:2018vsg,Obuljen:2018kdy}
\begin{equation}
    M_{\rm HI}(M) = M_0 \left( \frac{M}{M_{\rm min}}\right)^{\alpha}
    \exp\!\left[ - \left( \frac{M_{\rm min}}{M}\right)^{\beta} \right]\ ,
    \label{eq:HIHOD}
\end{equation}
with parameters $\thetaHOD=\{M_0,M_{\rm min},\alpha,\beta\}$.
This four-parameter form is intentionally simplified and is not meant to provide a fully realistic description of the post-reionization \HI--halo connection. In particular, it assumes that the \HI\ content depends only on halo mass, neglecting scatter at fixed mass, distinctions between centrals and satellites, possible assembly- or environment-dependent effects (known to be relevant for SBPs for galaxy surveys~\cite{Akitsu:2024lyt,Shiferaw:2024ehr}), and the extended spatial distribution of \HI\ within halos. Our purpose here is therefore not to advocate Eq.~\eqref{eq:HIHOD} as a complete physical model, but to use it as a controlled proof of concept for learning how variations in small-scale \HI\ occupation physics propagate into large-scale EFT bias parameters. More realistic prescriptions can be incorporated straightforwardly within the same framework by enlarging $\thetaHOD$ and modifying the \HI\ painting scheme.
Unless otherwise stated, we deposit the full \HI\ mass of each halo at the halo center.

We deposit \HI\ masses onto a Cartesian mesh using Cloud-In-Cell (CIC) assignment and define the \HI\ overdensity
\begin{equation}
\delta_{\rm HI}(\bm x)\equiv \frac{\rho_{\rm HI}(\bm x)}{\bar\rho_{\rm HI}}-1\ .
\label{eq:deltaHI_def}
\end{equation}
For HV, we use mesh choices that balance large-scale fidelity and computational cost; the fiducial examples shown below use $N_{\rm mesh}=256$. When working in Fourier space, we deconvolve the CIC window function for both $\delta_{\rm HI}$ and the operator fields.
Throughout this work, we scan the \HI--HOD parameters over the following ranges:
\begin{align*}
\log_{10}\!\left(\frac{M_0}{h^{-1}M_\odot}\right) &\in [9,11]\ , \\
\log_{10}\!\left(\frac{M_{\rm min}}{h^{-1}M_\odot}\right) &\in [5\times10^{10},\,12]\ , \\
\alpha &\in[-2,2]\ , \\
\beta &\in [0,2]\ .
\numberthis
\end{align*}
For comparison, the best-fit \HI--HOD parameters reported by Ref.~\cite{Villaescusa-Navarro:2018vsg} from analyses based on the \texttt{TNG300-1} simulations span roughly
$\alpha \simeq 0.24$--$0.79$,
$M_0 \simeq 1.4\times10^9$--$4.3\times10^{10}\,h^{-1}M_\odot$,
and
$M_{\rm min} \simeq 2\times10^{10}$--$2\times10^{12}\,h^{-1}M_\odot$
over $z=0$--$5$, while their best-fit models take $\beta=0.35$ at all redshifts.
The ranges adopted here are therefore intentionally broad enough to cover the expected region suggested by those fits (with the mild exception of $M_{\rm min}$), while also extending beyond it so that the learned HOD-to-bias mapping is not artificially restricted to a narrow neighborhood of previously calibrated models.

At each redshift, we generate 2000 \HI\ realisations by drawing
$\thetaHOD$ from this domain using Latin Hypercube
sampling~\cite{McKay:1979lhs}, which ensures efficient, space-filling
coverage of the four-dimensional HOD parameter space.
The same sampling ranges and strategy are adopted for both the
Hidden Valley and IllustrisTNG300 ensembles. 

Note that $\delta_{\rm HI}$ is normalized by the \emph{mean} \HI\ density in each realization. An overall rescaling of $M_{\rm HI}(M)$, controlled mainly by $M_0$, therefore largely cancels in the overdensity. Parameters that re-weight different halo masses---primarily $M_{\rm min}$ (via its appearance in the exponent) and $\alpha$, and to a lesser degree $\beta$---have a much stronger effect on clustering and on the inferred bias parameters, because they change the shape of the effective \HI\ weighting across the halo mass function rather than its normalization.

Further, the point-mass assignment and the neglect of any redistribution of gas among substructures within a halo are additional approximations. Their impact is expected to differ between our two simulation suites. For the Hidden Valley fields, which are constructed on a relatively coarse mesh with cell size $4.0\,h^{-1}{\rm Mpc}$, much of the sensitivity to intra-halo \HI\ structure should be pushed toward the grid scale and smaller scales. For \texttt{TNG300}, however, the finer mesh with cell size $0.8\,h^{-1}{\rm Mpc}$ makes the inferred field more sensitive to the small-scale \HI\ distribution within halos. We therefore reiterate that we regard the point-mass prescription primarily as a controlled proof-of-concept baseline---especially for the TNG comparison---rather than as a fully realistic description of the \HI\ field.

\subsection{Perturbative forward model}\label{sec:theory_shifted_ops}

The field-level approach models $\delta_{\rm HI}$ as a linear combination of \emph{shifted operators} built from the linear density field. The central idea is to treat large-scale advection non-perturbatively by shifting Lagrangian operators using the Zel'dovich displacement field, capturing IR displacements at the map level \cite{Schmittfull:2018yuk,Schmittfull:2020trd}. Full details of the application to \HI\ are contained in Ref.~\cite{Obuljen:2022cjo}, but we briefly summarize the approach here.

For each simulation realization, we acquire the linear density field $\delta_1(\bm k,z)$ at the same initial phases as the target \HI\ field. The first-order (Zel'dovich) displacement field is
\begin{equation}
\bm\Psi_1(\bm k,z)= i\,\frac{\bm k}{k^2}\,\delta_1(\bm k,z)\ .
\label{eq:psi1_def}
\end{equation}
In practice, $\delta_1$ and $\bm\Psi_1$ are evaluated on a regular Lagrangian grid and used in a particle-mesh procedure to generate shifted fields on the Eulerian grid.

If $\bm q$ denotes Lagrangian coordinates and $O(\bm q)$ denotes any operator constructed from $\delta_1(\bm q)$ and its derivatives,
a shifted (Eulerian) version of the operator can be written as
\begin{equation}
\widetilde{O}(\bm k)\equiv \int d^3q\;
O(\bm q)\,
\exp\!\left[-i\bm k\cdot\big(\bm q+\bm\Psi_1(\bm q)\big)\right]\ .
\label{eq:shifted_operator_def}
\end{equation}
We implement Eq.~\eqref{eq:shifted_operator_def} with a particle method: initialize particles at $\bm q$, evaluate $O(\bm q)$ and $\bm\Psi_1(\bm q)$, displace to $\bm x=\bm q+\bm\Psi_1(\bm q)$, and deposit particle weights $O(\bm q)$ onto the Eulerian mesh with CIC to obtain $\widetilde{O}(\bm x)$.

We adopt an operator set sufficient to describe deterministic \HI\ fluctuations through cubic order:
\begin{align}
O_1(\bm q) &= \delta_1(\bm q)\ , \nonumber\\
O_{\delta^2}(\bm q) &= \delta_1^2(\bm q)-\langle\delta_1^2\rangle\ , \nonumber\\
O_{\mathcal{G}_2}(\bm q) &= \mathcal{G}_2(\bm q)\ , \nonumber\\
O_{\delta^3}(\bm q) &= \delta_1^3(\bm q)-3\langle\delta_1^2\rangle\,\delta_1(\bm q)\ ,
\label{eq:operator_basis}
\end{align}
where the tidal operator $\mathcal{G}_2$ is given by
\begin{equation}
\mathcal{G}_2 \equiv
\left(\partial_i\partial_j \Phi_1\right)\left(\partial_i\partial_j \Phi_1\right)
-\frac{1}{3}\left(\nabla^2\Phi_1\right)^2
\label{eq:G2_def}
\end{equation}
with $\nabla^2\Phi_1=\delta_1$. 
At cubic order, this basis is intentionally truncated: we include only the local operator $O_{\delta^3}$ rather than the full set of independent third-order operators. In practice, this means that the fitted transfer function associated with $O_{\delta^3}$ should be interpreted as an \emph{effective} cubic response, since it can absorb contributions from omitted third-order terms in the same way as in the shifted-operator analyses of Refs.~\cite{Schmittfull:2018yuk,Obuljen:2022cjo}. This point is important when interpreting the inferred parameter $b_3$ below: it is the large-scale limit of the cubic coefficient within our chosen truncated basis, rather than a direct measurement of a unique physical cubic-bias parameter.

Following Ref.~\cite{Obuljen:2022cjo}, we decompose the simulated \HI\ field into a deterministic component captured by the operator model and a residual stochastic/error component:
\begin{equation}
\delta_{\rm HI}(\bm k)=\delta_{\rm HI,det}(\bm k)+\epsilon(\bm k)\ .
\label{eq:delta_split}
\end{equation}
The deterministic component is written as
\begin{equation}
\delta_{\rm HI,det}(\bm k)=\sum_a \beta_a(k)\,\widetilde{O}_a(\bm k)\ ,
\quad a\in\{1,\delta^2,\mathcal{G}_2,\delta^3\}\ ,
\label{eq:det_model_raw}
\end{equation}
where the $\beta_a(k)$ are isotropic scale-dependent transfer functions. The residual field $\epsilon$ contains stochasticity and any deterministic structure not captured by the finite operator basis. Its power spectrum,
\begin{equation}
P_{\rm err}(k)\equiv \langle|\epsilon(\bm k)|^2\rangle'\ ,
\label{eq:Perr_def}
\end{equation}
where the prime denotes removal of the momentum-conserving Dirac delta function and associated factors of $2\pi$, is a compact diagnostic of model performance.

\subsection{Estimation of transfer functions and bias parameters}\label{sec:theory_tf_estimation}

In practice, the shifted operators are correlated, so we work in an orthogonalized basis following Ref.~\cite{Foreman:2024kzw}. In that basis, transfer-function estimators reduce to ratios of cross spectra,
\begin{equation}
\widehat{\beta}_a(k)=
\frac{\langle \widetilde{O}_a^\perp(\bm k)\,\delta_{\rm HI}(-\bm k)\rangle'}
{\langle \widetilde{O}_a^\perp(\bm k)\,\widetilde{O}_a^\perp(-\bm k)\rangle'}\ .
\label{eq:beta_estimator}
\end{equation}
A practical advantage is that the operator fields and $\delta_{\rm HI}$ share the same initial phases, which yields strong sample-variance cancellation in the cross spectra and enables precise measurements even from a single simulation volume (as emphasized in Ref.~\cite{Obuljen:2022cjo}).

On sufficiently large scales, the transfer functions approach constants set by the bias parameters, while mild $k$-dependence appears from higher-derivative effects and other deterministic terms not included in the model. We extract $(b_1,b_2,b_{\mathcal G_2},b_3)$ by fitting smooth functional forms to the measured $\widehat{\beta}_a(k)$ on quasi-linear scales, following Ref.~\cite{Foreman:2024kzw}.
We fit
\begin{align}
\widehat{\beta}_1(k) &\approx b_1 + c_{1,1}\,k + c_{1,2}\,k^2 + c_{1,4}\,k^4\ , \label{eq:beta_fit_lin}\\
\widehat{\beta}_a(k) &\approx b_a + c_{a,2}\,k^2 + c_{a,4}\,k^4\ ,
\quad a\in\{\delta^2,\mathcal{G}_2,\delta^3\}\ , \label{eq:beta_fit_nonlin}
\end{align}
and identify the large-scale limits
\begin{equation}
b_1=\lim_{k\to 0}\widehat{\beta}_1(k)\ ,\quad
b_a=\lim_{k\to 0}\widehat{\beta}_{a}(k)\ .
\label{eq:bias_from_beta}
\end{equation}
Fits are performed in linearly spaced spherical $k$ bins. For the Hidden Valley box, the bin width is the fundamental mode,
\begin{equation}
\Delta k = \frac{2\pi}{L}
= 6.14\times10^{-3}\,h\,{\rm Mpc}^{-1}\,.
\end{equation}
For \texttt{TNG300}, the corresponding fundamental mode is larger because of the smaller box size,
$\Delta k=2\pi/L=3.06\times10^{-2}\,h\,{\rm Mpc}^{-1}$.
In these fits, each bin is weighted by the number of three-dimensional Fourier modes it contains.

\subsection{Examples of map-level reconstruction and transfer functions}\label{sec:hv_recon_tf}

Throughout this section, unless otherwise noted, we focus on $z=1$ as a representative redshift, since this is the regime where post-reionization 21\,cm measurements already exist
from CHIME and MeerKAT. The main trends are qualitatively similar across the full Hidden Valley redshift range, and the redshift evolution of the $b_i(b_1)$ relations, with $i=(2,\mathcal{G}_2,3)$, is shown in Appendix~\ref{app:b2_allz}.

\begin{figure*}
    \centering
    \includegraphics[width=1\linewidth]{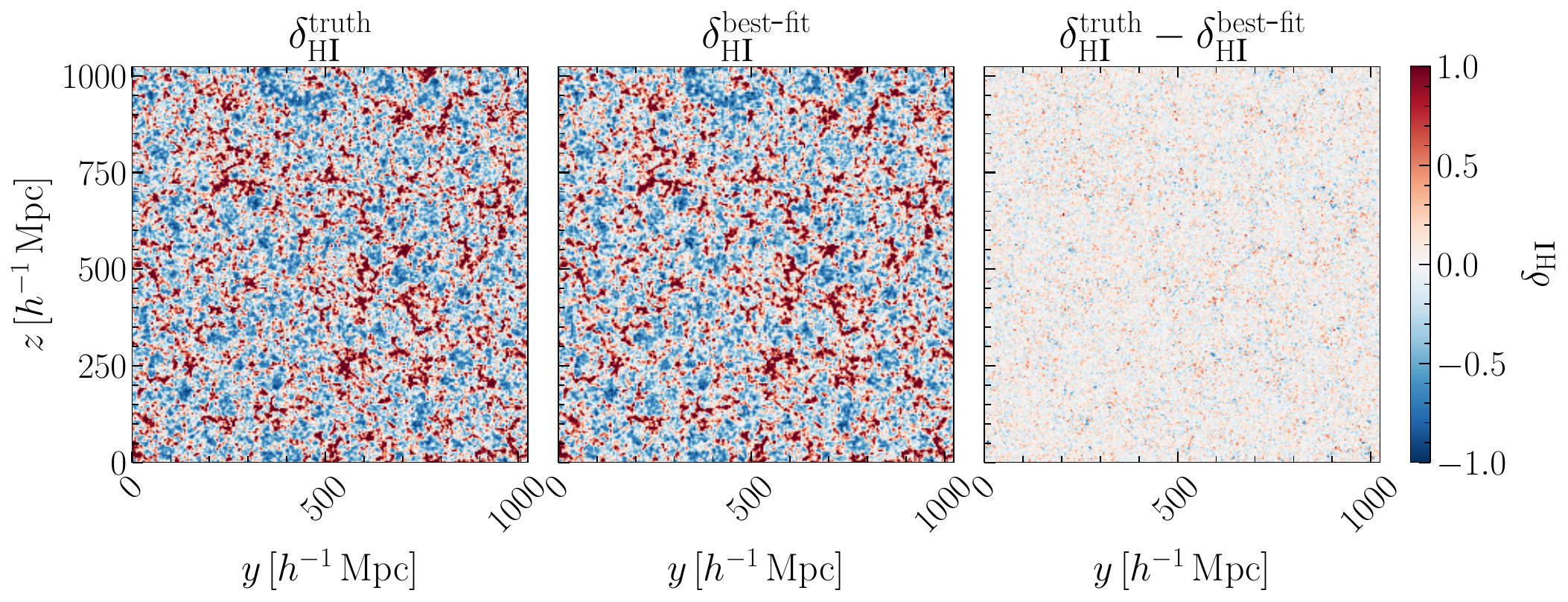}
    \caption{Hidden Valley field-level reconstruction of the \HI\ overdensity at $z=1$.
    Left: slice through the simulated truth \HI\ overdensity, $\delta_{\rm HI}^{\rm truth}$.
    Middle: best-fit deterministic reconstruction, $\delta_{\rm HI}^{\rm best\text{-}fit}$, obtained by fitting transfer functions in the shifted-operator basis.
    Right: residual field, $\delta_{\rm HI}^{\rm truth}-\delta_{\rm HI}^{\rm best\text{-}fit}$, shown on the same color scale.
    The close morphological agreement between the left and middle panels and the absence of coherent large-scale structure in the residual indicate that the deterministic model captures the dominant large-scale \HI\ fluctuations, leaving mainly stochastic and small-scale contributions in the residual.}
\label{fig:field_recons}
\end{figure*}

As a demonstration that the deterministic field-level model reproduces the phase-coherent large-scale structure of the \HI\ field,  Figure~\ref{fig:field_recons} shows a representative slice through a Hidden Valley \HI\ map from the $z=1$ snapshot, the best-fit reconstruction from the model, and the residual.

The truth map shows the standard cosmic-web morphology, and the reconstruction displays the same morphology, with individual large-scale features at the expected locations. 
The residual panel carries no obvious coherent large-scale structure aligned with the truth map---it is consistent with spatially uncorrelated noise. This is expected within the shifted-operator approach: once large-scale advection is absorbed into the relevant operators, the remaining mismatch should be dominated by stochasticity and small-scale physics not included in the deterministic basis. That behavior is what underpins the precision of the transfer-function estimates from a single volume: any residual large-scale coherence would appear as a systematic modeling error rather than noise, and would bias the low-$k$ plateau estimates that we use to extract the bias parameters.

\begin{figure}
    \centering
    \includegraphics[width=1\linewidth]{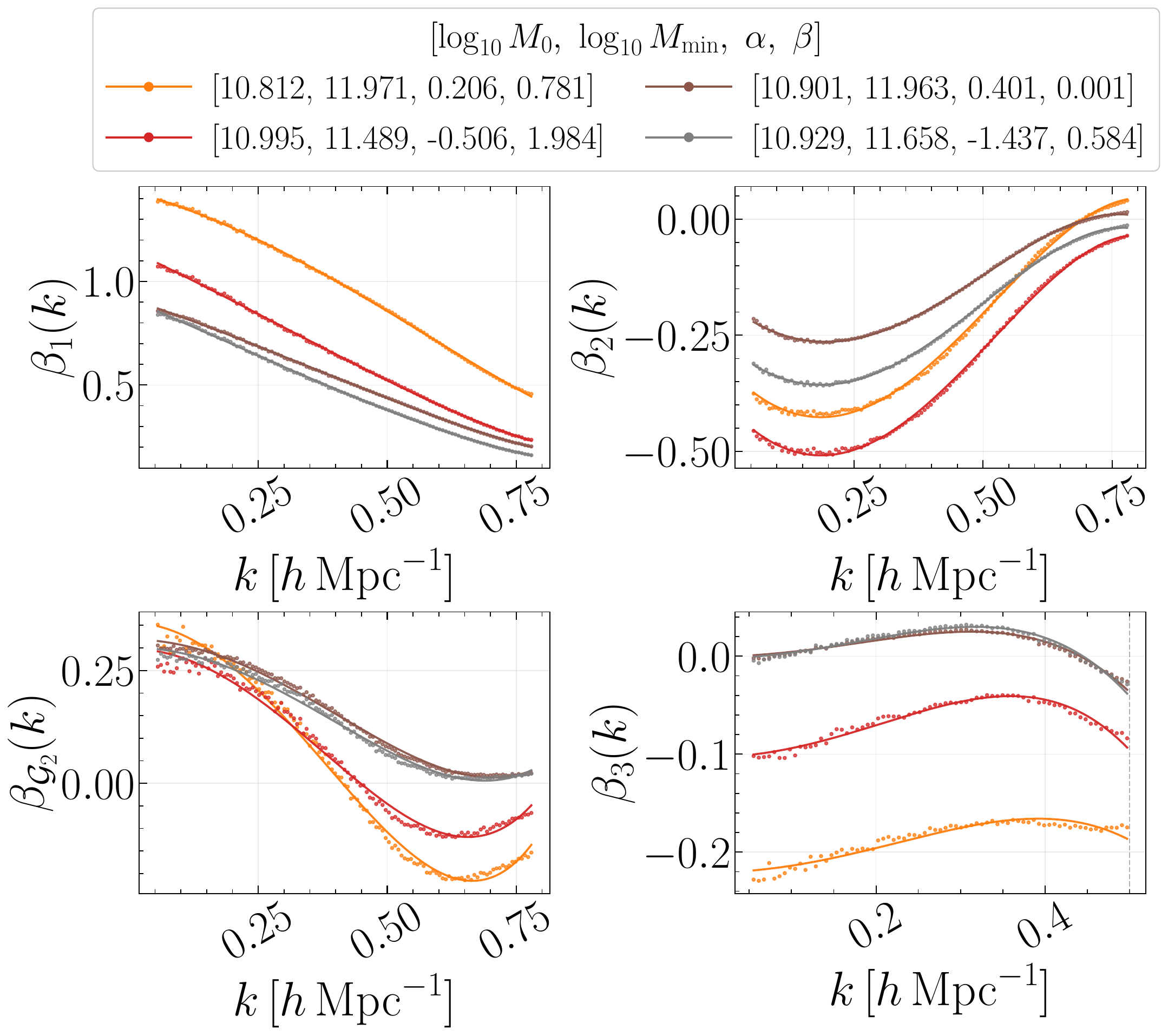}
    \caption{Hidden Valley transfer-function measurements at $z=1$ for four representative \HI--HOD choices.
    The legend reports the values of 
    $[\log_{10}(M_0/h^{-1}M_\odot),\log_{10}(M_{\rm min}/h^{-1}M_\odot),\alpha,\beta]$.
    Each panel shows the transfer function associated with one orthogonalized shifted operator: $\beta_1(k)$ (linear), $\beta_{\delta^2}(k)$ (quadratic local), $\beta_{\mathcal G_2}(k)$ (quadratic tidal), and $\beta_{\delta^3}(k)$ (cubic local).
    Points are binned estimates from Eq.~\eqref{eq:beta_estimator}, and the solid curves show the fits of Eqs.~\eqref{eq:beta_fit_lin}--\eqref{eq:beta_fit_nonlin} used to define $(b_1,b_2,b_{\mathcal G_2},b_3)$ through their $k\to0$ limits.
    While some of the transfer functions do not show a perfectly flat low-$k$ plateau over the plotted range, their large-scale behavior is smooth and well described by the fitting forms.
    The figure also shows that the response to HOD changes is modest for $\beta_1(k)$ and larger for the quadratic, tidal, and cubic operators.
    }
\label{fig:tf_example}
\end{figure}

Figure~\ref{fig:tf_example} shows measured transfer functions and their fits for several representative choices of \HI--HOD parameters.
In the plotted range, the transfer functions do not always exhibit a perfectly flat low-$k$ plateau, especially for $\beta_1(k)$, $\beta_{\mathcal G_2}(k)$, and $\beta_{\delta^3}(k)$.
For this reason, we do not define the bias parameters from any single low-$k$ bin.
Instead, we fit the smooth forms in Eqs.~\eqref{eq:beta_fit_lin}--\eqref{eq:beta_fit_nonlin} over the quasi-linear range and take their $k\to0$ limits.
This gives a stable way to extract the large-scale bias parameters even when some residual scale dependence remains across the fitted range.

The figure also shows that different operators respond differently to changes in the HOD parameters.
The variation in $\beta_1(k)$ across the four examples is visible but smaller than for the higher-order transfer functions.
The strongest changes appear in the quadratic, tidal, and cubic terms.
This is physically reasonable: changing the HOD reweights the \HI\ content across halo mass, which affects not only the overall clustering amplitude but also the nonlinear response of the tracer.
The tidal term is additionally sensitive to the anisotropic large-scale environment, so it is natural that $\beta_{\mathcal G_2}(k)$ shows a comparatively strong response.
This is qualitatively similar to the galaxy case explored in recent simulation-based-prior work~\cite{Ivanov:2024hgq,Ivanov:2024xgb,DESI:2025wzd,Chen:2025jnr,Chudaykin:2026nls}, where higher-order bias parameters are correlated with, but not fixed by, the linear bias.
Our \HI\ results suggest the same logic applies here, with additional structure tied to the underlying HOD parameters, so a dedicated \HI-specific simulation-based prior is well motivated.

\section{Relationship between bias and HOD parameters}\label{sec:hv-results}

\subsection{HOD-to-bias mapping}\label{sec:hv_hod_bias}

\begin{figure*}
    \centering
    \includegraphics[width=1\linewidth]{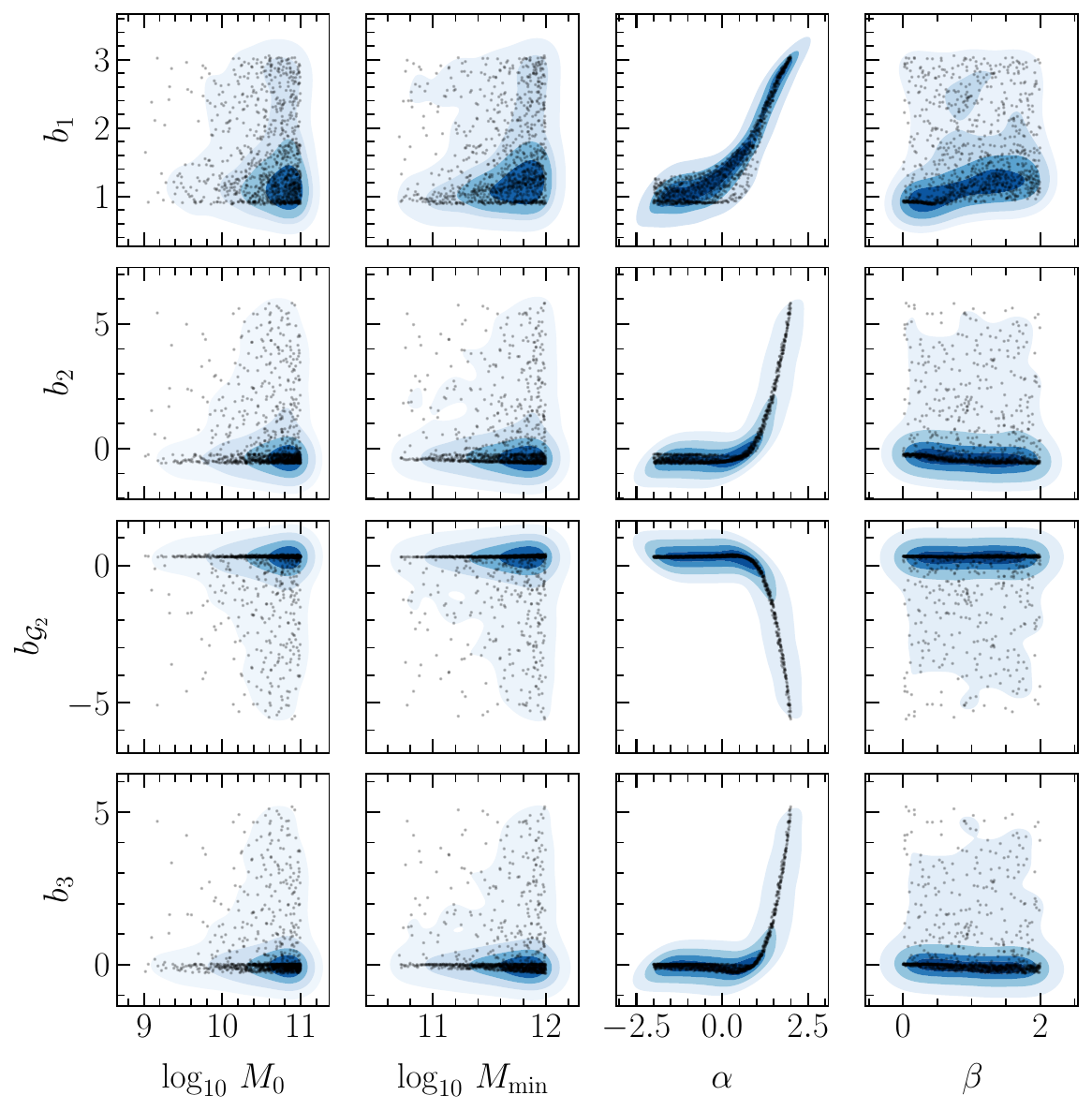}
    \caption{Hidden Valley: joint distributions between \HI--HOD parameters and inferred EFT bias parameters at $z=1$.
    Columns show the HOD parameters $[\log_{10}(M_0/h^{-1}M_\odot),\log_{10}(M_{\rm min}/h^{-1}M_\odot),\alpha,\beta]$ and rows show the inferred biases $(b_1,b_2,b_{\mathcal G_2},b_3)$.
    Gray points indicate individual \HI\ realizations in the HOD scan; blue contours show kernel-density estimates in each projection.
    The clearest dependence is on the high-mass slope $\alpha$, which strongly organizes all four bias parameters.
    The parameter $M_{\rm min}$ produces a visible trend mainly in $b_1$, while its effect on the higher-order biases is weaker in these one-dimensional projections.
    The parameters $M_0$ and $\beta$ show comparatively weak marginal trends.
    }
\label{fig:hod_bias_joint}
\end{figure*}

In this section, we examine the full empirical mapping from $\thetaHOD$ to $\thetaEFT=(b_1,b_2,b_{\mathcal{G}_2},b_3)$ across the sampled domain. Figure~\ref{fig:hod_bias_joint} summarizes the joint dependence of each bias parameter on each HOD parameter.

The $\alpha$ column shows the most pronounced trends. Increasing the high-mass slope $\alpha$ moves the ensemble toward larger $b_1$, $b_2$, and $b_3$, while $b_{\mathcal G_2}$ shifts to more negative values. The response is clearly nonlinear: both the local higher-order biases and the tidal bias follow curved trends as $\alpha$ is varied. This behavior is expected because increasing $\alpha$ gives more weight to massive halos in a nonlinear fashion, and these halos are more strongly biased and occupy different large-scale environments.

The $\log_{10}M_{\rm min}$ column shows a somewhat clearer trend for $b_1$ than for the higher-order parameters. Raising $M_{\rm min}$ suppresses the contribution from low-mass halos and shifts the effective \HI\ weight toward more massive halos, increasing the large-scale clustering amplitude. The corresponding trends in $b_2$, $b_{\mathcal G_2}$, and $b_3$ are weaker in this marginal projection. This indicates that $M_{\rm min}$ alone does not strongly organize the higher-order sector; its effect is partly degenerate with the other HOD parameters, especially $\alpha$.

The role of $\beta$ is also relatively weak in this projection. Larger $\beta$ sharpens the low-mass cutoff in Eq.~\eqref{eq:HIHOD}, which can shift the effective \HI\ weight upward in mass, but the induced changes in the bias parameters are modest compared to the changes driven by $\alpha$. Marginalizing over $\beta$ therefore mainly broadens the occupied region in bias space rather than setting its dominant structure.

The $\log_{10}M_0$ and $\log_{10}M_{\rm min}$ columns show broadly similar marginal structure in these projections. This is useful to keep in mind when interpreting the role of $M_{\rm min}$. While $M_0$ only controls the overall normalization of the \HI--HOD relation, $M_{\rm min}$ enters both the normalization factor $(M/M_{\rm min})^\alpha$ and the low-mass cutoff $\exp\!\left[ - \left( M_{\rm min}/M \right)^{\beta} \right]$. Since $\delta_{\rm HI}$ is normalized by the mean \HI\ density in each realization, a pure change in $M_0$ should cancel to leading order. The similarity of the two columns suggests that the visible $M_{\rm min}$ dependence is not simply an isolated cutoff effect, but is partly tied to how $M_{\rm min}$ changes the effective mass weighting and normalization of the HOD.

The main message from Fig.~\ref{fig:hod_bias_joint} is that the HOD-to-bias mapping is not captured by a set of independent Gaussian distributions. The $\alpha$ projections show strong curvature, and the widths of the bias distributions vary across the sampled HOD domain. At the same time, some parameters, such as $M_{\rm min}$ and $M_0$, are difficult to interpret from one-dimensional projections alone because their marginal trends can be weak or degenerate. This motivates the use of a conditional normalizing flow in Sec.~\ref{sec:flows}, which can learn the full multivariate distribution rather than relying on simple one-parameter trends.

\begin{figure}
    \centering
    \includegraphics[width=1\linewidth]{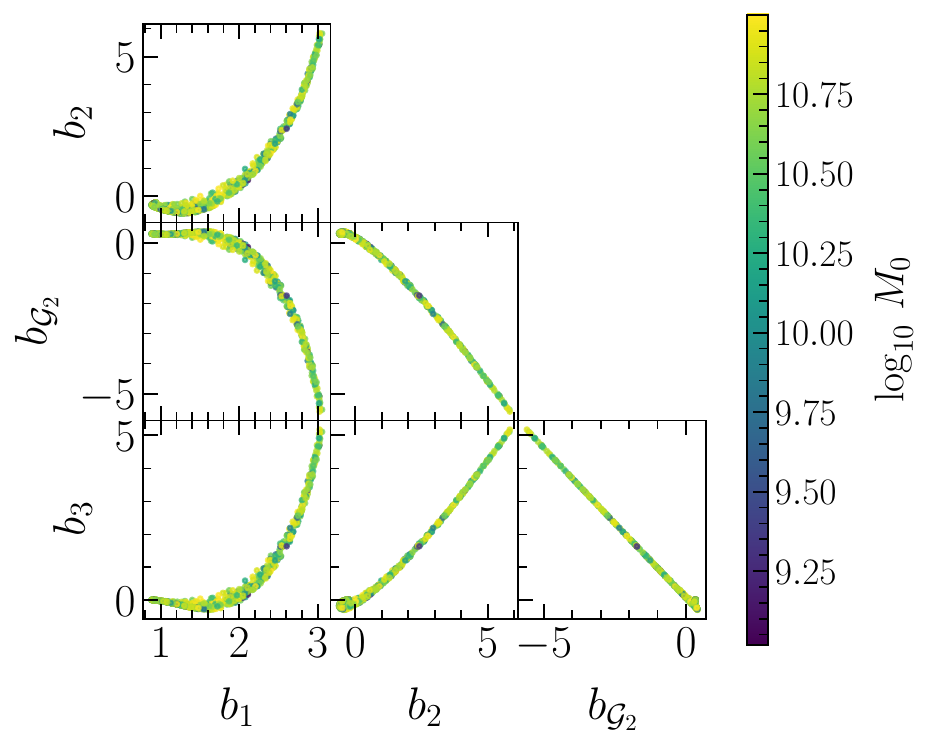}
    \caption{Hidden Valley: pairwise relations among inferred bias parameters at $z=1$, color-coded by $\log_{10}M_0$.
    Each point corresponds to one \HI\ realization in the HOD scan.
    The bias parameters occupy a narrow, curved manifold in $(b_1,b_2,b_{\mathcal G_2},b_3)$ space: $b_2$ and $b_3$ increase with $b_1$, while $b_{\mathcal G_2}$ becomes more negative at larger $b_1$.
    The color distribution is largely mixed, indicating that $M_0$ does not organize the motion along this manifold.}
\label{fig:bias_triangle_M0}
\end{figure}

\begin{figure}
    \centering
    \includegraphics[width=1\linewidth]{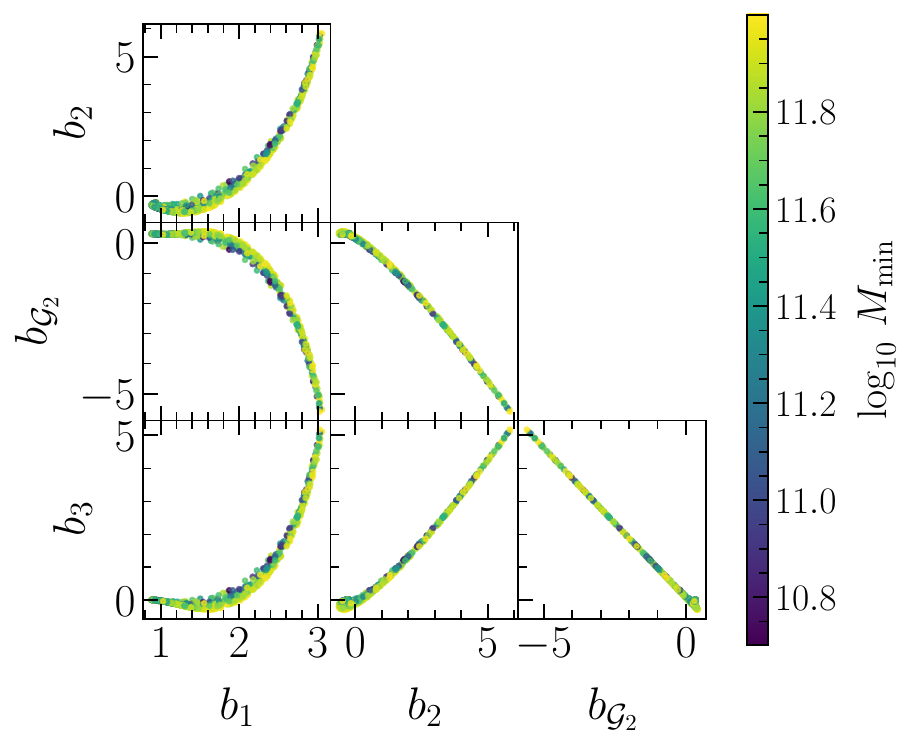}
    \caption{Hidden Valley: same bias--bias projections as Fig.~\ref{fig:bias_triangle_M0}, color-coded by $\log_{10}M_{\rm min}$. We find a weak trend that varying $M_{\rm min}$ changes the position of points across the thickness of the bias manifold, without moving them along the manifold's main direction.}
\label{fig:bias_triangle_Mmin}
\end{figure}

\begin{figure}
    \centering
    \includegraphics[width=1\linewidth]{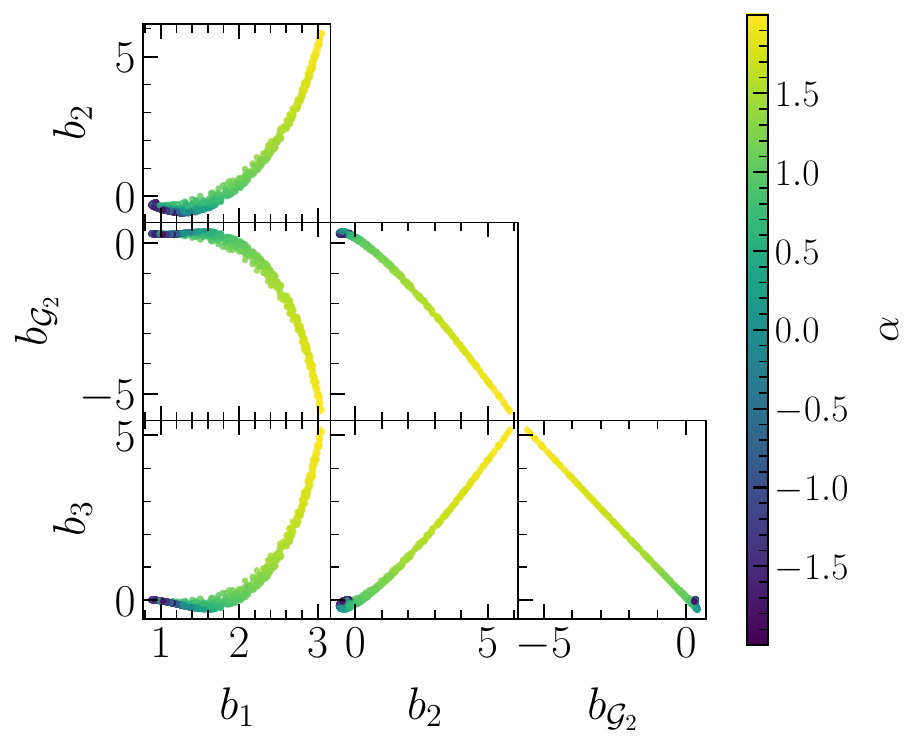}
    \caption{Hidden Valley: same as Fig.~\ref{fig:bias_triangle_M0}, color-coded by the high-mass slope $\alpha$. This parameter is strongly correlated with all bias parameter values, as expected from its influence on the relative weighting of different halo masses in our HOD model.}
\label{fig:bias_triangle_alpha}
\end{figure}

\begin{figure}
    \centering
    \includegraphics[width=1\linewidth]{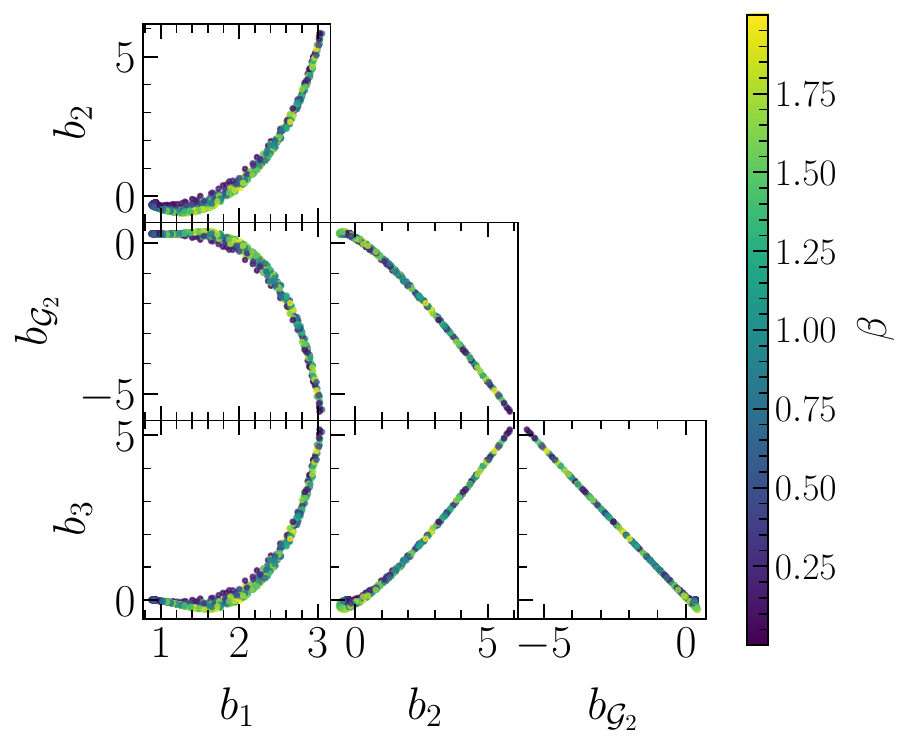}
    \caption{Hidden Valley: same as Fig.~\ref{fig:bias_triangle_M0}, color-coded by the cutoff-sharpness parameter $\beta$. The trend here is qualitatively similar to that of $M_{\rm min}$ in Fig.~\ref{fig:bias_triangle_Mmin}, with varying $\beta$ moving points orthogonally to the main direction of the bias manifold.}
\label{fig:bias_triangle_beta}
\end{figure}

One way to see which HOD parameters organize the occupied region in bias space is to color bias--bias projections by each HOD parameter in turn. Figures~\ref{fig:bias_triangle_M0}--\ref{fig:bias_triangle_beta} show the same set of bias pairings, with color indicating the relevant HOD parameter.

Figure~\ref{fig:bias_triangle_M0} reveals two key features. First, the bias parameters are far from independent: the ensemble occupies a narrow, curved region, and $b_2$ and $b_3$ rise steeply with $b_1$. 
This indicates that simulation-based priors incorporating these correlations are likely to improve the constraining power of full-shape analyses, compared to commonly-used uncorrelated priors.
Second, the colors are largely well-mixed along the main loci, indicating that $M_0$ is not organizing the distribution. This reinforces the picture that $M_0$ mostly sets the mean \HI\ density rather than the shape of the \HI\ overdensity field on large scales. Any remaining dependence on $M_0$ is better understood as an effect on stochasticity or on how much \HI\ is carried by rare objects, rather than a primary influence on deterministic large-scale response.

Figure~\ref{fig:bias_triangle_Mmin} shows a weak dependence on $M_{\rm min}$. At fixed or similar $b_1$, larger $M_{\rm min}$ tends to correspond to lower $b_2$ and $b_3$ and to less negative $b_{\mathcal G_2}$. This means that $M_{\rm min}$ changes the position of points across the thickness of the bias manifold, but does not move them along the main $b_i(b_1)$ trends.

The clearest color ordering is produced by the high-mass slope $\alpha$, shown in Fig.~\ref{fig:bias_triangle_alpha}. In most projections, increasing $\alpha$ moves realizations along the main bias manifold toward larger $b_1$, larger local higher-order biases $b_2$ and $b_3$, and more negative $b_{\mathcal G_2}$. This trend is physically expected: larger $\alpha$ gives greater weight to massive halos, which are more strongly clustered and therefore occupy the high-bias end of the ensemble. However, the upper panels show that much of the variation in the higher-order biases is already tightly correlated with $b_1$. Thus, Fig.~\ref{fig:bias_triangle_alpha} should not be interpreted as demonstrating an independent dependence on $\alpha$ at fixed $b_1$; rather, it shows that $\alpha$ is strongly correlated with the location of a realization along the same nonlinear bias--bias relations. This distinction is important for constructing simulation-based priors: even when the dominant structure can be summarized by relations involving $b_1$, the full conditional model remains useful because it preserves the joint, curved, and non-Gaussian support of $(b_1,b_2,b_{\mathcal G_2},b_3)$ generated by the HOD scan.

The $\beta$-colored projections in Fig.~\ref{fig:bias_triangle_beta} show a weaker ordering than the $\alpha$-colored case, but they are not entirely featureless. Points with different $\beta$ values are mixed over much of the full range in $b_1$, $b_2$, $b_{\mathcal G_2}$, and $b_3$, indicating that $\beta$ is not the dominant parameter controlling motion along the main bias manifold. At the same time, the upper panels show behavior qualitatively similar to the $M_{\rm min}$-colored projections: changing either parameter can move realizations up and down across the thickness of the $b_2(b_1)$, $b_{\mathcal G_2}(b_1)$, and $b_3(b_1)$ loci. This is physically reasonable because both $M_{\rm min}$ and $\beta$ control the low-mass suppression in Eq.~\eqref{eq:HIHOD}: $M_{\rm min}$ sets the characteristic scale of the cutoff, while $\beta$ controls how sharply that cutoff is applied. Thus, $\beta$ acts as a secondary HOD direction whose main effect is to broaden and reshape the conditional distribution around the dominant bias--bias relations.

\begin{figure*}
    \centering
    \includegraphics[width=1\linewidth]{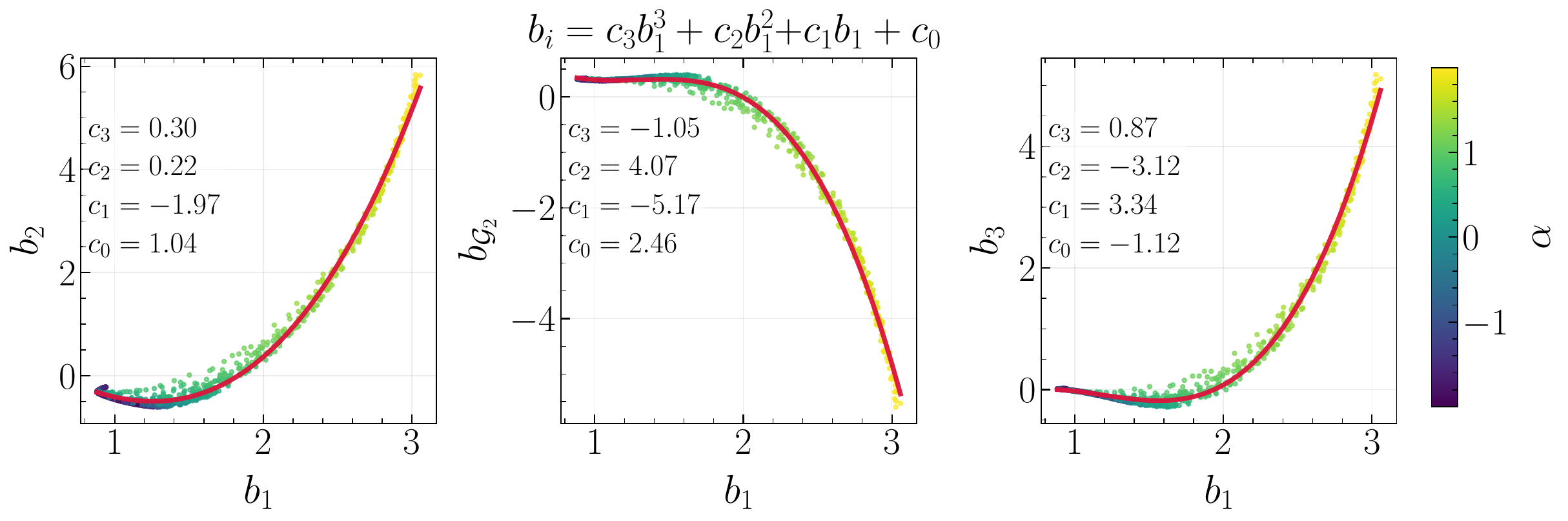}
    \caption{Hidden Valley: empirical bias--bias relations at $z=1$.
    Each panel shows a higher-order bias parameter as a function of the linear bias $b_1$: $b_2$ (left), $b_{\mathcal G_2}$ (middle), and $b_3$ (right).
    Points represent individual realizations from the HOD scan and are color-coded by $\alpha$.
    Red curves show cubic polynomial fits in $b_1$ as a visual guide to the mean trend.
    The relations are strongly nonlinear, and the color gradient shows that the high-mass slope $\alpha$ organizes the position of points along these relations.
    }
\label{fig:bias_vs_b1_poly_alpha}
\end{figure*}

In Figure~\ref{fig:bias_vs_b1_poly_alpha}, we show polynomial fits to each higher-order bias as a function of $b_1$, overlaid on the full distributions color-coded by $\alpha$.
These one-parameter relations are not meant to replace the full conditional mapping, but they summarize the curvature and scatter in a form that is easy to compare across simulations. 

All three relations are clearly nonlinear: the higher-order biases grow rapidly with $b_1$, and $b_{\mathcal G_2}$ becomes more negative at larger $b_1$. This is in line with the tight loci already visible in the previous plots.

\subsection{Comparison to analytic HI-mass-weighted halo expectations}\label{sec:bias_sim_vs_ana}

As a sanity check, we compare our field-level measurements to a simple analytic prediction that treats \HI\ as an \HI-mass-weighted halo tracer, by analogy with standard expressions for galaxy bias (e.g.~\cite{Baugh:1998cw,Benson:1999mva}). For each set of HOD parameter values $\thetaHOD$ at redshift $z$, the mean \HI\ density can be written in terms of the halo mass function $dn/dM$ and \HI-mass--halo-mass relation $M_{\rm HI}(M;\thetaHOD)$ as
\begin{equation}
\bar{\rho}_{\rm HI}(z)=\int dM\,\frac{dn}{dM}(M,z)\,M_{\rm HI}(M;\thetaHOD)\ .
\end{equation}
The \HI-weighted Eulerian bias parameters can likewise be written in terms of the $n$th-order halo bias $b_n^h(M,z)$ as 
\begin{widetext}
\begin{equation}
b_n^{\rm HI}(z)=\frac{1}{\bar{\rho}_{\rm HI}(z)}\int dM\,\frac{dn}{dM}(M,z)\,M_{\rm HI}(M;\thetaHOD)\,b_n^{h}(M,z)\ ,
\label{eq:hi_weighted_biases}
\end{equation}
\end{widetext}
where $n=1,2,3$.
We evaluate $dn/dM$ and $b_1^h(M,z)$ in the HV cosmology using \texttt{pyccl}\footnote{\url{https://github.com/LSSTDESC/CCL}}~\cite{LSSTDarkEnergyScience:2018yem} with fitting functions from Ref.~\cite{Tinker:2010my} (using the $200{\rm m}$ mass definition). For higher-order halo biases, we use the relations in Ref.~\cite{Lazeyras:2015lgp}, expressed as functions of $b_1^h$:
\begin{align}
b_2^h(b_1^h) &= 0.412 - 2.143\,b_1^h + 0.929\,(b_1^h)^2 + 0.008\,(b_1^h)^3\ , \label{eq:b2h_of_b1h}\\[4pt]
b_3^h(b_1^h) &= -1.028 + 7.646\,b_1^h - 6.227\,(b_1^h)^2 + 0.912\,(b_1^h)^3\ . \label{eq:b3h_of_b1h}
\end{align}

\begin{figure*}
    \centering
    \includegraphics[width=1\textwidth]{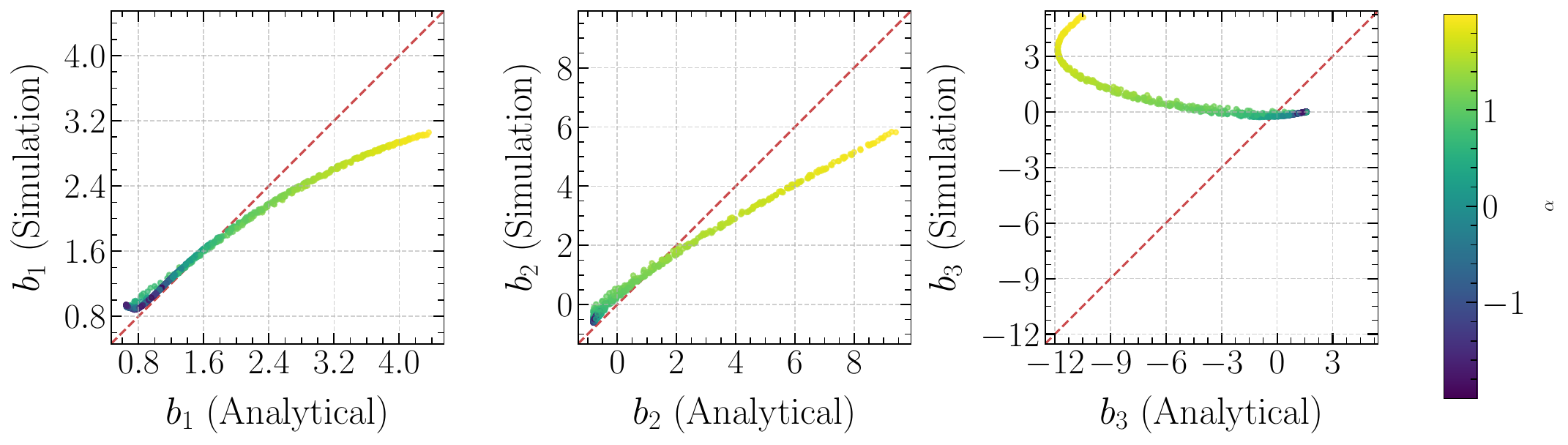}
    \caption{Hidden Valley: simulation-inferred versus analytic bias parameters at $z=1$.
Each panel shows a scatter plot of one bias parameter, with the 
$x$-axis giving the analytic \HI-mass-weighted halo prediction 
(Eqs.~\eqref{eq:hi_weighted_biases}--\eqref{eq:b3h_of_b1h}) and 
the $y$-axis giving the field-level measurement from transfer-function 
fits, for each \HI--HOD realisation in our ensemble.
Points are color-coded by $\alpha$; the dashed red line indicates $y=x$.
The deviations from the diagonal are structured and grow progressively 
from $b_1$ to $b_2$ to $b_3$, 
where the analytic prediction fails 
qualitatively for $\alpha>0$, indicating that a simple \HI-mass-weighted 
halo-bias picture is insufficient to reproduce the effective EFT 
parameters across the full HOD domain.}
\label{fig:bias_sim_vs_ana}
\end{figure*}

Figure~\ref{fig:bias_sim_vs_ana} compares the simulation-inferred bias 
parameters to the analytic \HI-mass-weighted halo predictions of 
Eqs.~\eqref{eq:hi_weighted_biases}--\eqref{eq:b3h_of_b1h}, for all 
HOD realisations at $z=1$. Each panel plots one bias parameter, with 
points color-coded by $\alpha$.

The analytic baseline captures the broad ordering of the field-level measurements, but it does not reproduce the inferred bias parameters in detail. For both $b_1$ and $b_2$, the simulation-inferred values broadly increase with the analytic prediction, indicating that the simple \HI-mass-weighted halo-bias picture captures part of the underlying mass-weighting trend. However, neither relation lies cleanly on the diagonal: both show systematic offsets and departures from a one-to-one mapping, with the discrepancy becoming more noticeable toward the high-bias end.
When considering potential causes for this discrepancy, note that \HI\ assembly bias is not included in the analytical expression from Eq.~\eqref{eq:hi_weighted_biases}, nor in our procedure for painting \HI\ mass into halos. However, our analytical expression also ignores \textit{halo} assembly bias, while our halo catalogs do include such effects (to the extent that they are captured by the underlying $N$-body simulations). Therefore, halo assembly bias may contribute to the discrepancy we observe.

The most significant discrepancy appears in the cubic sector. The 
simulation-inferred $b_3$ does not follow the analytic 
\HI-mass-weighted halo prediction, and the points form a strongly 
curved locus rather than scattering around the diagonal. This should 
not be interpreted only as a failure of the halo-bias fitting formula. 
In our field-level model, $b_3$ is the coefficient of the cubic local 
operator included in the chosen basis, and it can absorb contributions 
from other third-order operators that are not explicitly included in 
the model. The comparison in the $b_3$ panel is therefore a useful 
diagnostic of the effective cubic response measured by our pipeline, 
but it should not be read as a direct one-to-one test of the analytic 
halo $b_3^h$ relation, or the bias-halo relation in Eq.~\eqref{eq:hi_weighted_biases}.

The color coding shows that the deviations are organized by the HOD 
parameters rather than being random scatter. In particular, $\alpha$ is 
strongly correlated with the inferred bias parameters because it changes 
the relative weight assigned to massive halos. This reinforces the 
main point of the comparison: simple \HI-mass-weighted analytic 
relations provide a useful baseline, but they do not reproduce the full 
effective bias parameters measured from the fields. For this reason, 
we learn $p(\thetaEFT\mid\thetaHOD)$ directly from the simulation 
ensemble rather than imposing a fixed halo-calibrated functional form.

\section{Normalizing flows for simulation-based priors}\label{sec:flows}

The measurements described in the previous section yield paired samples
$\{(\thetaHOD^{(i)},\thetaEFT^{(i)})\}_{i=1}^{N}$, where
$\thetaEFT=(b_1,b_2,b_{\mathcal G_2},b_3)$ are inferred as described in Sec.~\ref{sec:theory}, and
$\thetaHOD=(\log_{10}(M_0/h^{-1}M_\odot),\log_{10}(M_{\rm min}/h^{-1}M_\odot),\alpha,\beta)$ specifies the \HI--HOD parameters used to generate each \HI\ field.
The goal is to learn a flexible model of the conditional density $p(\thetaEFT\mid\thetaHOD)$.
Once learned, this conditional distribution supports sampling-based SBPs and provides a compact way to visualize which directions in HOD parameter space control each bias parameter.

\subsection{Density estimation with normalizing flows}\label{sec:flow_basics}

Normalizing flows are likelihood-based generative models that represent a complicated target density as the pushforward of a simple base distribution through an invertible transformation~\cite{Kobyzev2019NormalizingFI, Papamakarios2019NormalizingFF, JimenezRezende2015VariationalIW, Papamakarios2019NormalizingFF, Kong2020TheEP}.
Let $u\sim\pi(u)$ be a base density, which we take to be a standard normal in $D$ dimensions, and let $f_{\varphi}$ be an invertible map such that
\begin{equation}
\theta = f_{\varphi}(u)\ .
\end{equation}
The induced density is then
\begin{equation}
\hat p_{\varphi}(\theta)
=
\pi\!\left(f_{\varphi}^{-1}(\theta)\right)\,
\left|\det\left(\frac{\partial f_{\varphi}^{-1}}{\partial \theta}\right)\right|\ .
\label{eq:flow_cov}
\end{equation}
Here $\varphi$ denotes the trainable weights and biases of the neural networks that parameterize the transformation.

In our case, the data at fixed redshift consist of paired samples
$\{(\thetaHOD^{(i)},\thetaEFT^{(i)})\}_{i=1}^{N}$,
with
$\thetaEFT=(b_1,b_2,b_{\mathcal G_2},b_3)$
and
$\thetaHOD=(\log_{10}(M_0/h^{-1}M_\odot),\log_{10}(M_{\rm min}/h^{-1}M_\odot),\alpha,\beta)$.

To learn the conditional density $p(\thetaEFT\mid\thetaHOD)$, we utilize a conditional flow, in which the invertible map depends on a context vector
$c\equiv\thetaHOD$:
\begin{align}
\nonumber
\hat p_{\varphi}(\thetaEFT\,|\,\thetaHOD)
&=
\pi\!\left(f_{\varphi}^{-1}(\thetaEFT;\thetaHOD)\right) \\
&\quad\times
\left|\det\left(\frac{\partial f_{\varphi}^{-1}}{\partial \thetaEFT}\right)\right|\ .
\label{eq:flow_conditional}
\end{align}
Operationally, this means that the neural networks inside the autoregressive transform take $\thetaHOD$ as an additional input, so different HOD points define different invertible maps and hence different densities in EFT space.

The trainable parameters $\varphi$ are determined by maximum likelihood. For the conditional model, we minimize the negative log-likelihood over a training set with size $N_{\rm tr}$,
\begin{equation}
\mathcal{L}(\varphi)
=
-\frac{1}{N_{\rm tr}}
\sum_{i\in{\rm tr}}
\log \hat p_{\varphi}\!\left(\thetaEFT^{(i)}\mid\thetaHOD^{(i)}\right)\ .
\label{eq:flow_nll}
\end{equation}
After training, the result is a fitted model
$\hat p_{\hat\varphi}(\thetaEFT\mid\thetaHOD)$
that can be evaluated at arbitrary HOD points and sampled directly.

We emphasize that the conditional character of Eq.~\eqref{eq:flow_conditional} arises because $\thetaHOD$ is treated as a context variable: the invertible transformation acts only on $\thetaEFT$, and the Jacobian determinant is therefore evaluated only with respect to $\thetaEFT$. A joint density $p(\thetaEFT,\thetaHOD)$ could instead be learned by training a flow on the concatenated eight-dimensional vector $(\thetaEFT,\thetaHOD)$. Alternatively, once a prior $p(\thetaHOD)$ is specified, the present conditional model defines the corresponding joint distribution through
\begin{equation}
p(\thetaEFT,\thetaHOD)
= p(\thetaEFT\mid\thetaHOD)\,p(\thetaHOD)\,.
\end{equation}

\subsection{Architecture, preprocessing, and training}\label{sec:flow_arch_preproc}

Before training, we standardize both the HOD parameters and the EFT parameters feature-by-feature using $z$-score normalization,
\begin{equation}
x \rightarrow \frac{x-\bar x}{\sigma_x}\ ,
\end{equation}
where $\bar x$ and $\sigma_x$ are the sample mean and standard deviation of each feature.
This choice is better matched to the Gaussian base distribution used by the flow, since it places each feature on a comparable scale without introducing hard boundaries. In practice, it avoids the density pile-up near the edges that can occur with min--max scaling when the transformed variables are modeled with Gaussian tails.
We store these quantities so that samples drawn from the trained flow can be mapped back to physical units.

We then randomly permute the list of paired samples and split it into 80\% training, 10\% validation, and 10\% test sets.
The random permutation has no physical meaning; it simply removes any accidental ordering in the stored sample list before the split is made.
The model is optimized only on the training set.
The validation set is used for model selection and early stopping, and the test set is held back entirely until the end.

We implement the flow with \texttt{nflows}\footnote{\url{https://github.com/bayesiains/nflows}}~\cite{nflows} and use a masked autoregressive neural spline architecture.
The conditional model for
$\hat p(\thetaEFT\mid\thetaHOD)$
is built from six masked piecewise rational-quadratic autoregressive transforms, each preceded by a random permutation of the features.
Each transform uses 128 hidden features, 8 spline bins, linear tails beyond a standardized tail bound of 5, residual blocks, GELU activations, and dropout probability 0.1.
This architecture is better suited to the strongly non-Gaussian conditional distributions seen in the EFT bias parameters.

For completeness, we also train an unconditional flow for
$\hat p(\thetaEFT)$,
but the main object used below is the conditional model.

We optimize the training loss in Eq.~\eqref{eq:flow_nll} with Adam using learning rate
$3\times10^{-4}$,
weight decay
$10^{-5}$,
and batch size 128.
We evaluate the validation loss every 250 gradient steps, apply gradient clipping with maximum norm 1.0, and use a cosine-annealing learning-rate schedule with a floor of $10^{-6}$.
Training is stopped early if the validation loss fails to improve for 20 consecutive evaluation intervals.
Whenever the validation loss improves, we save the corresponding model checkpoint, and after training we restore the checkpoint with the best validation loss.
The final fitted density
$\hat p_{\hat\varphi}(\thetaEFT\mid\thetaHOD)$
therefore corresponds to the best-validation model, not simply to the last gradient step.

\begin{figure}
    \centering
    \includegraphics[width=1\linewidth]{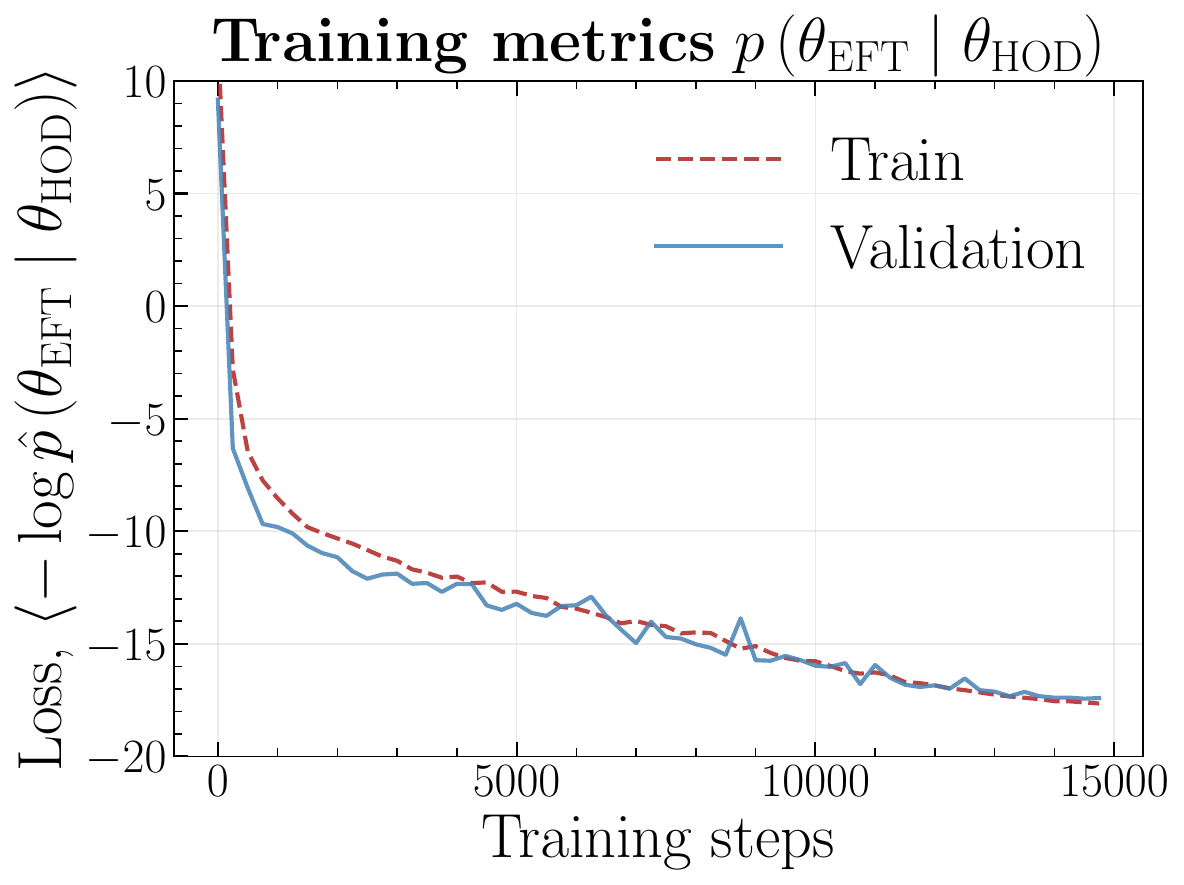}
    \caption{Training and validation negative log-likelihood,
    $-\langle\log\hat{p}(\thetaEFT\mid\thetaHOD)\rangle$,
    for the conditional normalizing flow at $z=1$.
    The model is optimized only on the training set; the validation curve is computed without gradient updates and is used for checkpointing and early stopping.
    Both curves decrease rapidly at early times and then track each other closely until convergence, indicating that the fitted flow generalizes well to unseen data.
    }
    \label{fig:loss}
\end{figure}

Figure~\ref{fig:loss} shows the evolution of the training and validation negative log-likelihood during optimization.
The rapid drop in both curves at early times and their close tracking near convergence indicate that the model is fitting the training data without a strong loss of performance on the validation split.
A separate 10\% test set is not shown in the figure and is used only after training as a fully held-out check on the final model.

In what follows, we use two main diagnostics for the conditional flow: the loss curves in Fig.~\ref{fig:loss} and one-at-a-time conditional slices through the trained model.

\subsection{Conditional-flow behavior}\label{sec:hv_flow_results}

A direct way to inspect what the flow has learned is to vary one HOD parameter at a time while holding the other three fixed to their sample means.
Because the HOD parameters are standardized before training, fixing the remaining three to their means corresponds to setting them to zero in the standardized space.
At each point along the resulting one-dimensional slice, we draw conditional samples from the flow, transform them back to physical units, and summarize the resulting bias distribution by its median and its 16th and 84th percentiles.

\begin{figure*}
    \centering
    \includegraphics[width=1\textwidth]{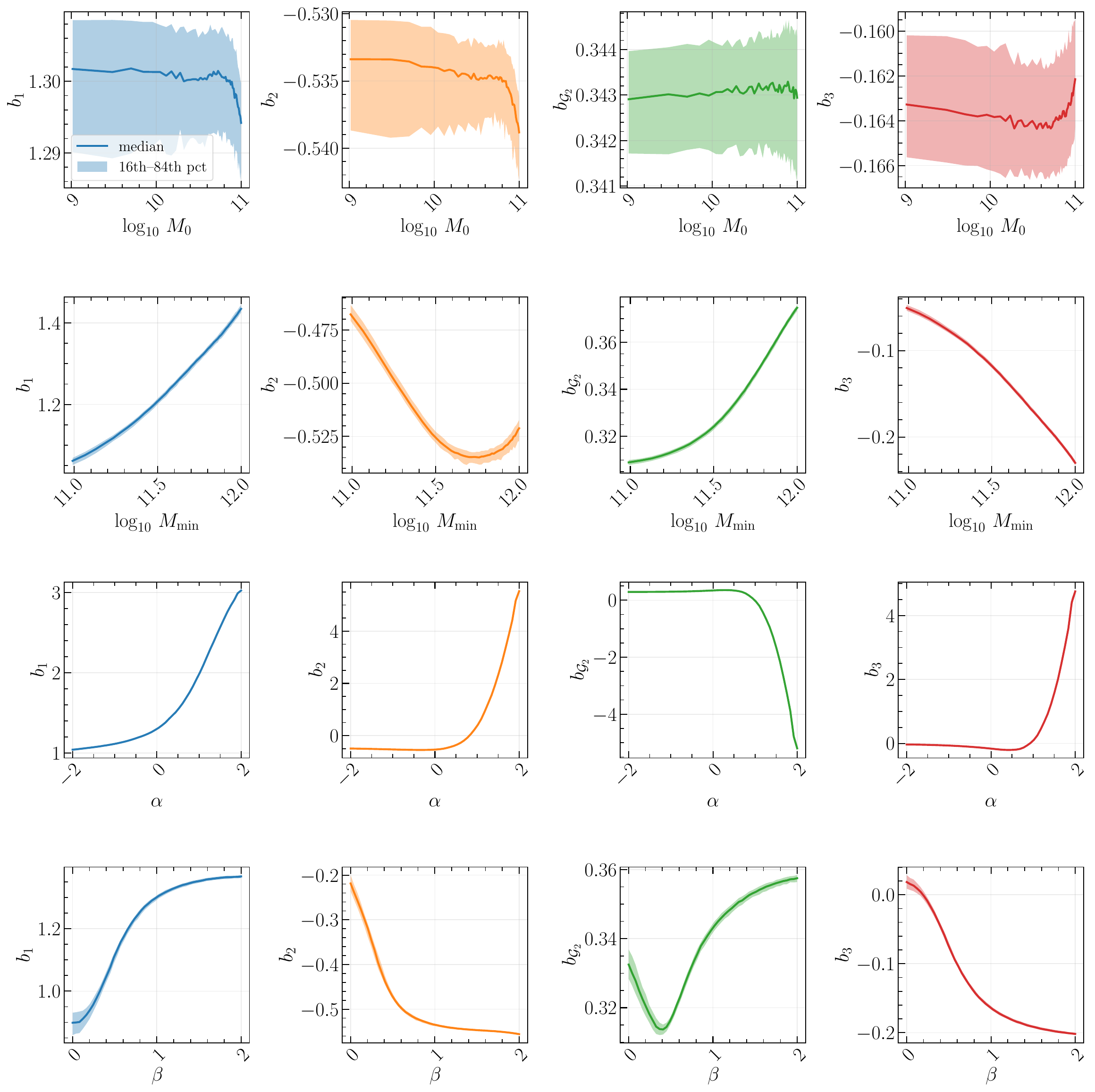}
    \caption{Hidden Valley: one-at-a-time conditional slices through the trained normalizing flow at $z=1$.
    Each row varies one HOD parameter across its physical training range while holding the other three fixed to their sample means.
    Columns show the four inferred EFT bias parameters $(b_1,b_2,b_{\mathcal G_2},b_3)$.
    Solid curves show the conditional median predicted by the flow,
    representing the mean response of each bias parameter to each HOD parameter, and the shaded bands show the 16th--84th percentile range of the conditional distribution.
    Note that, these curves should not be interpreted as reconstructions of the
    two-dimensional projections in Fig.~\ref{fig:hod_bias_joint}.
    In that figure, the other three HOD parameters vary across the full
    ensemble and are therefore marginalized over in each projection,
    whereas here they are fixed to their sample means. The two figures
    thus show marginal and conditional views of the HOD-to-bias mapping,
    respectively.
    }
\label{fig:flow_1d_sweeps}
\end{figure*}

Figure~\ref{fig:flow_1d_sweeps} shows that the flow has learned a strongly anisotropic HOD-to-bias mapping.
The weakest dependence is on $M_0$: all four medians are nearly flat as $\log_{10}M_0$ is varied, consistent with the fact that the overall \HI\ normalization largely cancels in the overdensity field.
By contrast, varying $M_{\rm min}$ produces clear and coherent shifts in all four biases.
Across the slice shown here, increasing $M_{\rm min}$ raises $b_1$ and $b_{\mathcal G_2}$, while pushing $b_2$ and $b_{3}$ to more negative values.

The strongest response is seen when varying the high-mass slope $\alpha$.
Positive $\alpha$ rapidly increases $b_1$, $b_2$, and $b_3$, while driving $b_{\mathcal G_2}$ to large negative values.
This matches the physical picture already seen in the earlier sections: increasing $\alpha$ upweights massive halos and therefore changes both the overall clustering amplitude and the higher-order bias sector.

The dependence on $\beta$ is weaker than the dependence on $\alpha$, but it is not negligible.
Its effect is partly monotonic for $b_1$, $b_2$, and $b_3$, and more clearly non-monotonic for $b_{\mathcal G_2}$.
This is physically reasonable, since both $M_{\rm min}$ and $\beta$ control the suppression of low-mass halos in Eq.~\eqref{eq:HIHOD}, so changing one can partially mimic changing the other.
The conditional response of the flow therefore separates a strong direction, mainly associated with $\alpha$, from weaker and partly degenerate directions associated with $M_{\rm min}$ and $\beta$.

The width of the 16th--84th percentile bands also carries information.
Where the bands are narrow, the map from HOD space to EFT space is close to deterministic along that slice.
Where they broaden, different combinations of the remaining HOD parameters can lead to similar conditional medians.
This is most obvious in the nearly flat $M_0$ slices, where the mean response is small but the conditional uncertainty remains finite.
This is precisely the information that is useful for SBP applications: the trained flow returns tight conditional priors where the HOD-to-bias mapping is sharp, and broader ones where degeneracies remain.

\section{Comparison of IllustrisTNG and Hidden Valley}\label{sec:tng}

The results above rely on Hidden Valley as the primary training ground for the HOD-to-bias mapping in this paper. As a cross-check, we now repeat a subset of the same measurements on IllustrisTNG300 halo catalogs using the same \HI--HOD parameterization and the same field-level bias-inference pipeline, then compare the resulting trends to Hidden Valley.

\subsection{HOD-to-bias projections and bias relations}

\begin{figure*}
    \centering
    \includegraphics[width=\linewidth]{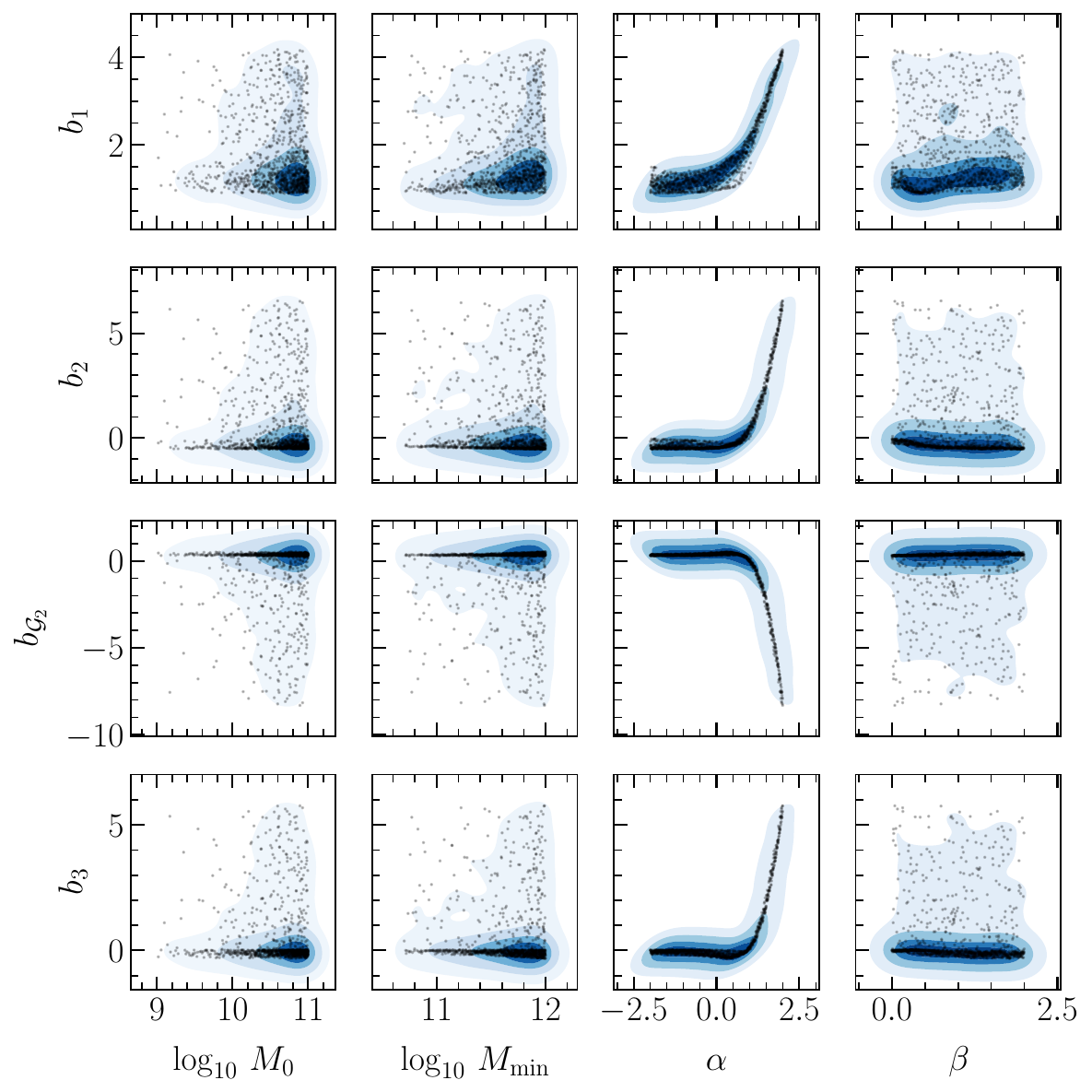}
    \caption{IllustrisTNG300: joint dependence of inferred bias parameters on \HI--HOD parameters at $z=1$, with the same layout as Fig.~\ref{fig:hod_bias_joint}.
    Points show individual HOD realizations, and contours show kernel-density estimates in each projection.
    Relative to Hidden Valley, the occupied region extends to larger $b_1$, $b_2$, and $b_3$, and especially to more negative $b_{\mathcal G_2}$.
    As in Hidden Valley, the strongest organization is with $\alpha$, while the trends with $M_0$ and $\beta$ remain weak.
    }
\label{fig:tng_hod_bias_pairs}
\end{figure*}

\begin{figure*}
    \centering
    \includegraphics[width=\textwidth]{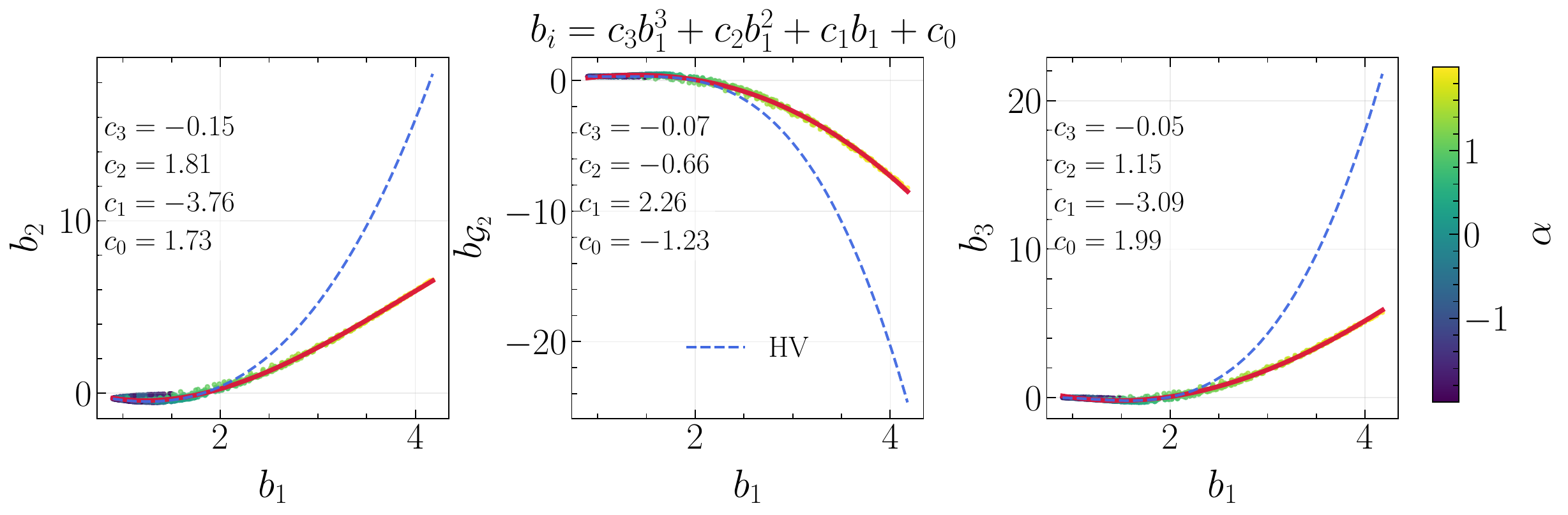}
    \caption{IllustrisTNG300: bias relations as functions of $b_1$ at $z=1$, color-coded by $\alpha$ (as in Fig.~\ref{fig:bias_vs_b1_poly_alpha}).
    The higher-order biases follow tight one-parameter trends with $b_1$, with clear curvature at low and intermediate $b_1$ and a strong ordering with $\alpha$ along the relations. The dashed blue curves show the best-fit cubic polynomials from the Hidden Valley simulation (Fig.~\ref{fig:bias_vs_b1_poly_alpha}).
    }
\label{fig:tng_bias_relations_alpha}
\end{figure*}

Figure~\ref{fig:tng_hod_bias_pairs} shows the joint dependence of $(b_1,b_2,b_{\mathcal G_2},b_3)$ on the HOD parameters in \texttt{TNG300}, in the same format as Fig.~\ref{fig:hod_bias_joint}. Figure~\ref{fig:tng_bias_relations_alpha} summarizes the corresponding bias relations as functions of $b_1$.

The \texttt{TNG300} projections show the same broad qualitative pattern seen in Hidden Valley: the mapping is nonlinear, the occupied region in bias space is narrow compared to the full four-dimensional space, and $\alpha$ is the clearest organizing direction. At the same time, the differences from Hidden Valley are not small in all sectors. In particular, \texttt{TNG300} reaches larger values of $b_1$, $b_2$, and $b_3$, and its tidal bias extends to substantially more negative values. These shifts are most pronounced in the high-bias part of the distribution.

For the intended application to simulation-based priors, this means that the two simulations agree on the overall structure of the mapping but not on its detailed support everywhere in parameter space. That is encouraging at the level of qualitative behavior, but it also suggests that the choice of simulation may matter quantitatively when constructing precision priors, especially in the high-bias tail and in the tidal sector.

The bias--bias relations in \texttt{TNG300} (Fig.~\ref{fig:tng_bias_relations_alpha}) are again tight functions of $b_1$. The $b_2(b_1)$ and $b_3(b_1)$ relations are clearly curved at low and intermediate $b_1$, and then steepen toward the high-bias end. The tidal bias $b_{\mathcal G_2}$ decreases rapidly with $b_1$ and spans a much wider range than in Hidden Valley. As in the joint projections, the color coding shows that $\alpha$ is strongly correlated with position along these relations.

\begin{figure*}
\centering
\includegraphics[width=\linewidth]{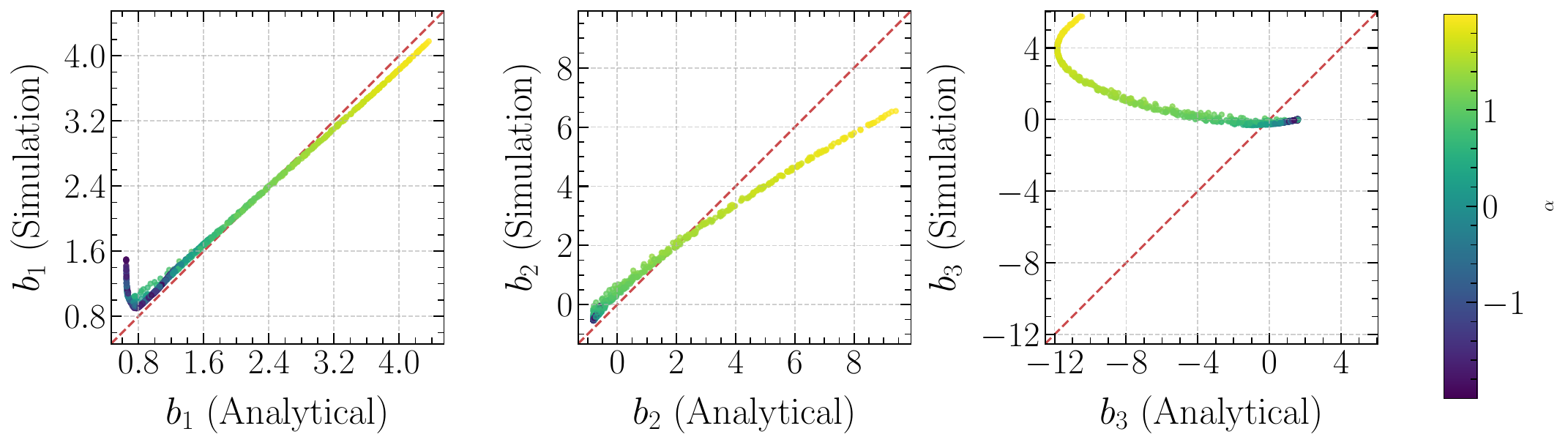}
\caption{IllustrisTNG300: simulation-inferred versus analytic bias parameters at $z=1$, in the same format as Fig.~\ref{fig:bias_sim_vs_ana}. Points are color-coded by the high-mass slope $\alpha$, and the dashed red line indicates equality. Relative to the Hidden Valley comparison, the agreement is noticeably better for $b_1$ and $b_2$, although $b_2$ still shows a systematic offset at the high-bias end. The cubic response $b_3$ remains poorly described by the analytic prediction.}
\label{fig:bias_sim_vs_ana_tng}
\end{figure*}

As an additional comparison, Fig.~\ref{fig:bias_sim_vs_ana_tng} repeats the analytic \HI-mass-weighted halo-bias test of Fig.~\ref{fig:bias_sim_vs_ana} for \texttt{TNG300}. The same analytic prescription is used, with the halo mass function and halo-bias relations evaluated in the \texttt{TNG300} cosmology.

Compared with the Hidden Valley result in Fig.~\ref{fig:bias_sim_vs_ana}, the \texttt{TNG300} measurements are closer to the analytic prediction in the $b_1$ and $b_2$ panels. The linear bias follows the one-to-one relation over most of the dynamic range, apart from the low-bias branch at small or negative $\alpha$. The quadratic bias also preserves the analytic ordering and lies closer to the diagonal than in Hidden Valley, although the simulation-inferred values fall below the analytic prediction at the largest $b_2$. The cubic sector remains qualitatively different: the field-level $b_3$ values form a curved locus that is far from the one-to-one relation, especially for large positive $\alpha$. Thus, the improved agreement in \texttt{TNG300} is mainly a feature of the lower-order local biases, while the effective cubic response still requires a simulation-calibrated treatment.

\subsection{Conditional-flow sweeps in \texttt{TNG300}}

To further compare the learned HOD-to-bias mapping in the two simulation suites, we repeat the same one-at-a-time sweep diagnostic from Fig.~\ref{fig:flow_1d_sweeps} for a flow trained on the \texttt{TNG300}-derived pairs, and overplot the corresponding Hidden Valley median relations.

\begin{figure*}[t]
    \centering
    \includegraphics[width=\textwidth]{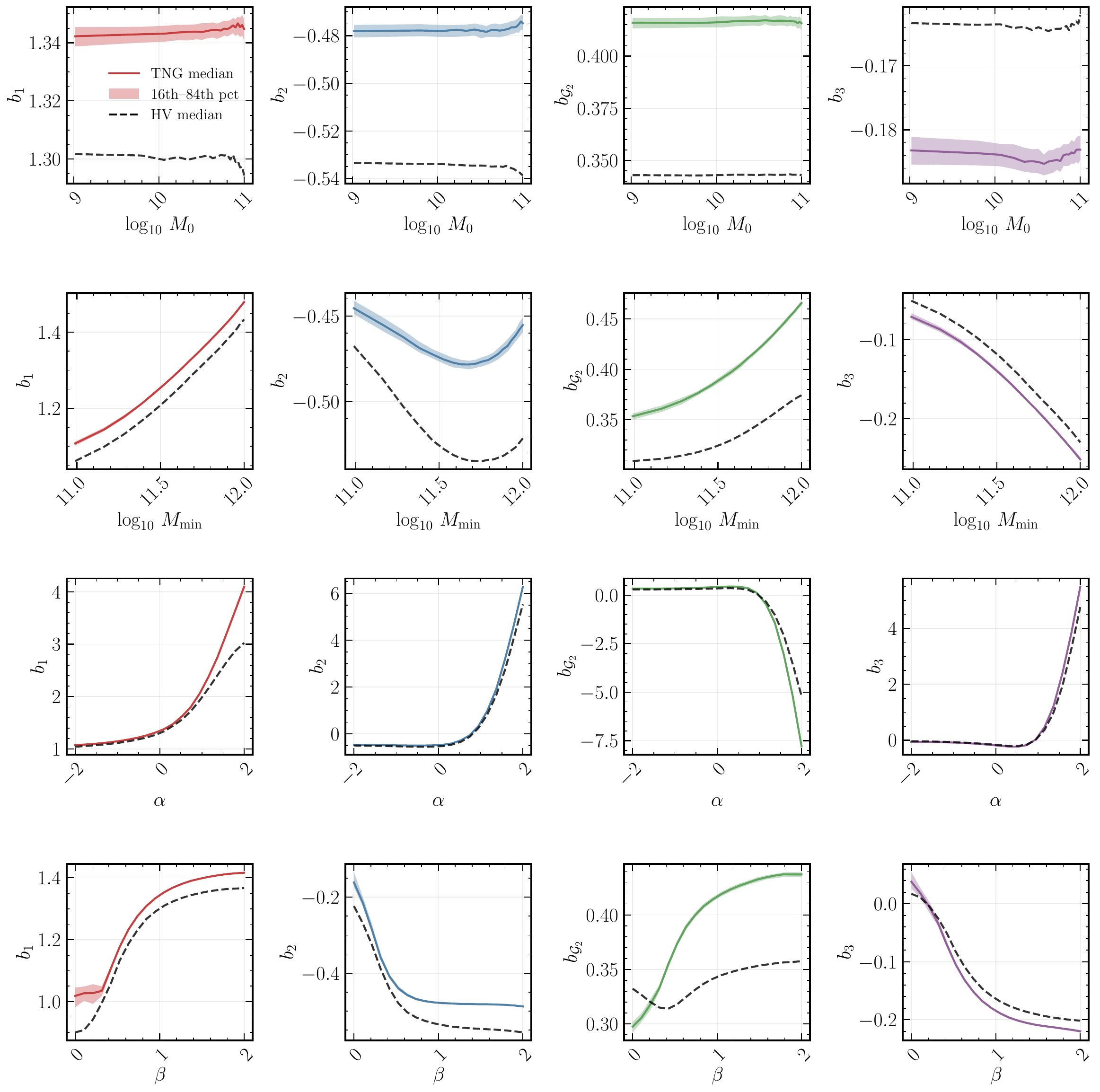}
    \caption{IllustrisTNG300: one-at-a-time conditional slices through the trained normalizing flow, compared directly to the Hidden Valley medians.
    In each row, one HOD parameter is varied across its sampled range while the other three are held fixed to their mean values.
    Solid curves show the \texttt{TNG300} conditional medians and shaded bands show the 16th--84th percentile range; dashed curves show the corresponding Hidden Valley medians.
    The two simulations show the same broad hierarchy of parameter sensitivities, but with systematic offsets in several panels, most notably in $b_{\mathcal G_2}$ and, more modestly, in $b_2$ and $b_3$ for the $M_{\rm min}$ and $\beta$ sweeps.}
\label{fig:tng_flow_sensitivity}
\end{figure*}

The TNG sweeps in Fig.~\ref{fig:tng_flow_sensitivity} preserve the same broad hierarchy of sensitivities seen in Hidden Valley: varying $M_0$ produces the weakest response, varying $\alpha$ produces the strongest response, and $M_{\rm min}$ and $\beta$ give intermediate trends. The main difference between the two flows is therefore not which HOD directions matter, but the normalization of the resulting bias relations.

The closest agreement appears in the $\alpha$ sweeps, where the TNG and Hidden Valley medians are similar in $b_1$, $b_2$, and $b_3$, with somewhat larger differences only in $b_{\mathcal G_2}$ at the high-$\alpha$ end. Larger offsets appear in the $M_{\rm min}$ and $\beta$ sweeps. In particular, \texttt{TNG300} predicts systematically larger $b_{\mathcal G_2}$ at fixed $M_{\rm min}$ and fixed $\beta$, and it also shows smaller but visible shifts in $b_2$ and $b_3$ relative to Hidden Valley. The $M_0$ sweeps are nearly flat in both simulations, but they are offset from each other by an approximately constant amount.

The 16th--84th percentile bands are generally narrow and do not vary strongly across most of the sweeps, so the dominant effect is a systematic difference between the TNG-based and Hidden-Valley-based median mappings. For broad or conservative simulation-based priors, these shifts may still be less important than the overall gain relative to completely uninformative priors. For precision applications, however, the offsets are large enough to matter, especially in the tidal sector.

\subsection{Halo-mass dependence of the HV--TNG bias 
comparison}\label{sec:hv_tng_masssplit}

\begin{figure*}
    \centering
    \includegraphics[width=\linewidth]{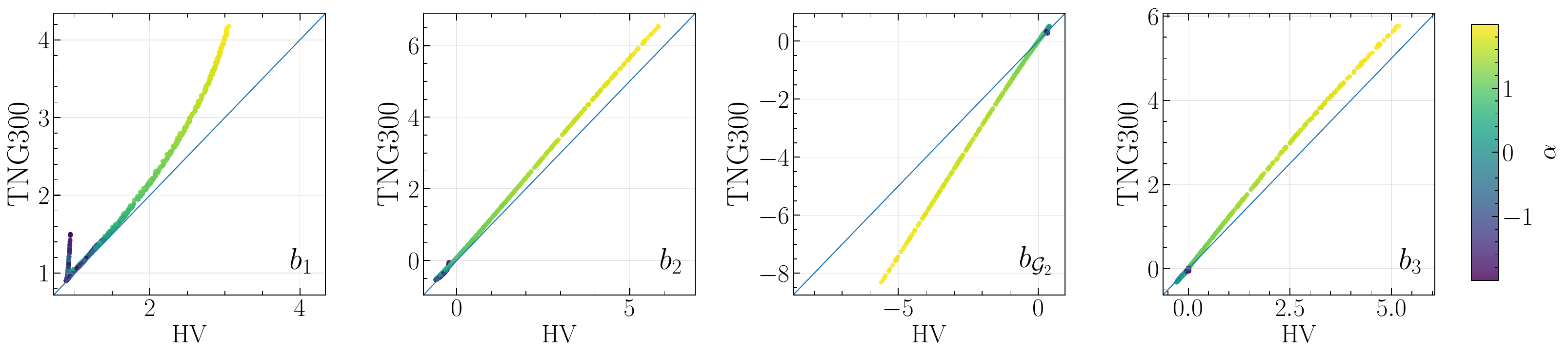}
    \caption{Hidden Valley vs.\ IllustrisTNG300: comparison of inferred bias parameters at $z=1$ for the full matched HOD sample.
    Each panel plots one bias parameter with HV on the $x$-axis and \texttt{TNG300} on the $y$-axis, for matched HOD realizations color-coded by $\alpha$.
    The diagonal blue line indicates perfect agreement.
    For $b_1$, $b_2$, and $b_3$, most points lie above the diagonal, indicating larger values in \texttt{TNG300} than in HV.
    For $b_{\mathcal G_2}$, the offset has the opposite sign: \texttt{TNG300} is generally less negative than HV for the same HOD input.}
    \label{fig:hv_tng_comp_full}
\end{figure*}

\begin{figure*}
    \centering
    \includegraphics[width=\linewidth]{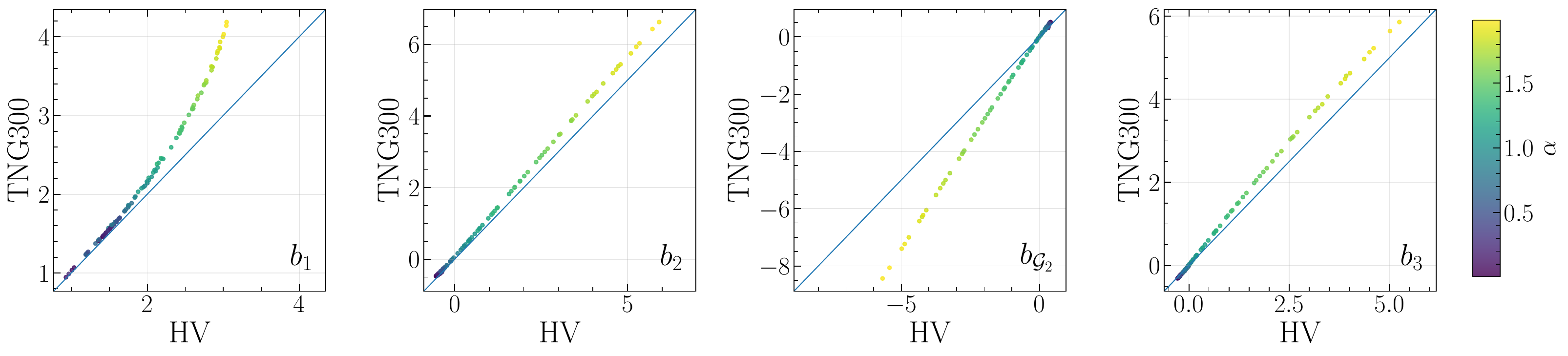}
    \caption{Same as Fig.~\ref{fig:hv_tng_comp_full}, for halos with $M_h>10^{10}\,h^{-1}M_\odot$ and $\alpha>0$.
    This selection retains the broad high-bias structure of the full comparison.
    \texttt{TNG300} remains systematically above HV in $b_1$, $b_2$, and $b_3$, while $b_{\mathcal G_2}$ remains systematically less negative than in HV.}
    \label{fig:hv_tng_comp_gt10}
\end{figure*}

\begin{figure*}
    \centering
    \includegraphics[width=\linewidth]{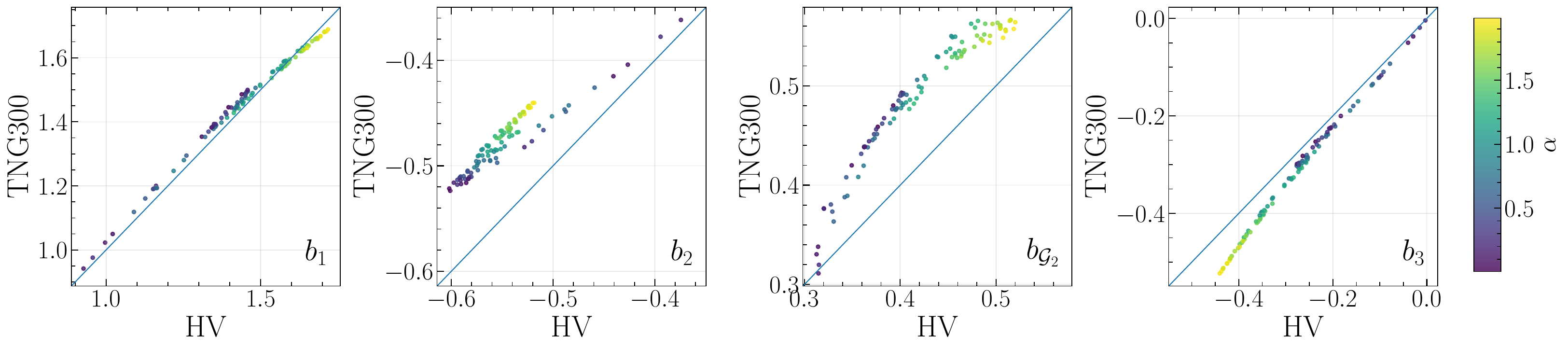}
    \caption{Same as Fig.~\ref{fig:hv_tng_comp_full}, but restricted to halos with $M_h\in[10^{10},10^{13}]\,h^{-1}M_\odot$ and $\alpha>0$.
    Relative to the full sample, the dynamic range shrinks substantially, especially in $b_{\mathcal G_2}$ and $b_3$, showing that the highest-mass halos strongly affect the comparison.
    The two simulations show good agreement for $b_1$, while the other three parameters remain discrepant.
    }
    \label{fig:hv_tng_comp_10_13}
\end{figure*}

\begin{figure*}
    \centering
    \includegraphics[width=\linewidth]{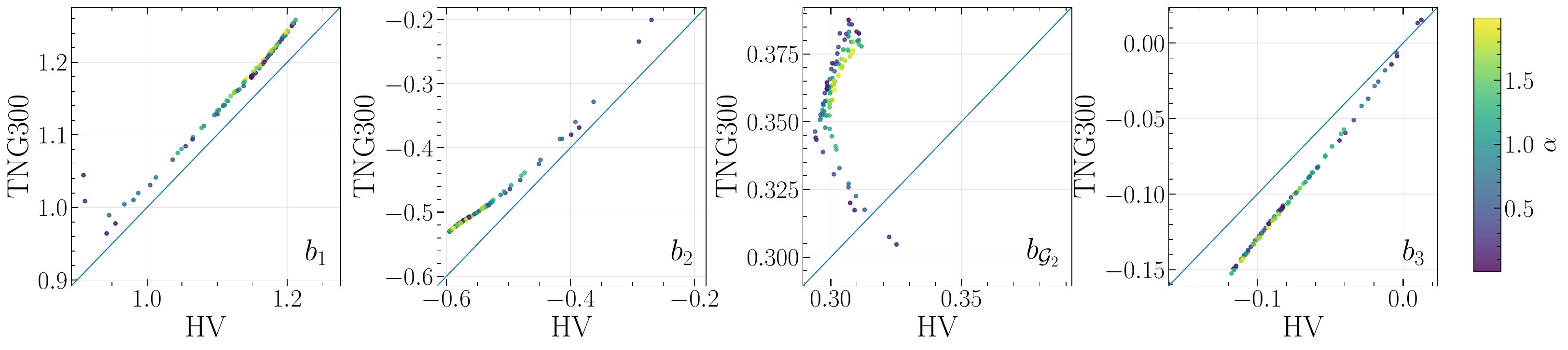}
    \caption{Same as Fig.~\ref{fig:hv_tng_comp_full}, restricted to halos with $M_h<10^{12}\,h^{-1}M_\odot$ and $\alpha>0$.
    Removing the high-mass tail shifts both simulations into a different bias regime: $b_{\mathcal G_2}$ becomes positive and $b_3$ becomes negative.
    However, the two simulations do not collapse onto the diagonal; noticeable offsets remain in all four parameters.}
    \label{fig:hv_tng_comp_lt12}
\end{figure*}

The direct comparison between Hidden Valley and IllustrisTNG300 in Fig.~\ref{fig:hv_tng_comp_full} shows that, for matched HOD realizations, the inferred bias parameters are not identical between the two simulations. 
To investigate whether the offsets between the two are driven mainly by particular halo-mass ranges, we construct matched \HI\ fields from both simulations using 100 HOD realizations with $\alpha>0$ and other parameters spanning the ranges
$M_0\in[10^9,10^{11}]\,h^{-1}M_\odot$,
$M_{\rm min}\in[5\times10^{10},10^{12}]\,h^{-1}M_\odot$,
and $\beta\in[0,2]$,
and compare the four inferred bias parameters at $z=1$ for four halo selections: the full sample, halos with $M_h\in[10^{10},10^{13}]\,h^{-1}M_\odot$, halos with $M_h>10^{10}\,h^{-1}M_\odot$, and halos with $M_h<10^{12}\,h^{-1}M_\odot$.
The results are shown in Figs.~\ref{fig:hv_tng_comp_full}--\ref{fig:hv_tng_comp_lt12}, with each point representing one HOD realization colored by $\alpha$. The diagonal blue line marks perfect agreement.

The full-sample comparison (Fig.~\ref{fig:hv_tng_comp_full}) shows a clear, parameter-dependent pattern. For $b_1$, $b_2$, and $b_3$, \texttt{TNG300} generally yields larger values than Hidden Valley for the same HOD input, and the offset grows toward larger $\alpha$. The effect is strongest in the high-bias part of the sample. The tidal sector also shows a large discrepancy: \texttt{TNG300} generally gives \emph{more negative} values of $b_{\mathcal G_2}$ than HV. Thus the two simulations differ not only in amplitude, but also in how the same HOD maps into the higher-order and tidal bias sectors.

Comparing the restricted samples clarifies which halo masses matter most. The $M_h>10^{10}\,h^{-1}M_\odot$ sample (Fig.~\ref{fig:hv_tng_comp_gt10}) looks broadly similar to the full sample, which shows that simply removing the very lowest-mass halos does not erase the inter-simulation differences. By contrast, imposing an upper cut at $10^{13}\,h^{-1}M_\odot$ (Fig.~\ref{fig:hv_tng_comp_10_13}) changes the comparison substantially, especially for $b_{\mathcal G_2}$, although it now produces similar $b_1$ values between the two simulations. This indicates that the most massive halos play an important role in setting the extreme tail of the bias mapping.

Restricting further to $M_h<10^{12}\,h^{-1}M_\odot$ (Fig.~\ref{fig:hv_tng_comp_lt12}) changes the inferred bias regime even more strongly. In this sample, both simulations move to smaller $b_1$, negative $b_2$, positive $b_{\mathcal G_2}$, and negative $b_3$. This confirms that the high-mass halo population has a major effect on the sign and amplitude of the higher-order bias parameters. However, this cut does \emph{not} produce close agreement between the two simulations. \texttt{TNG300} still tends to lie above HV in $b_1$, $b_2$, and $b_{\mathcal G_2}$, while the $b_3$ comparison remains offset and scattered.  
The main lesson is therefore not that the low-mass-restricted sample solves the discrepancy, but that different halo-mass ranges probe qualitatively different parts of the HOD-to-bias mapping.

Several physical mechanisms could potentially  contribute to the offsets observed in 
the full-mass comparison. \texttt{TNG300} includes baryonic processes 
--- AGN feedback, stellar winds, gas stripping --- that modify 
halo masses, concentrations, and shapes relative to the 
dark-matter-only HV runs, particularly at 
$M_h\gtrsim10^{12}\,h^{-1}M_\odot$ where AGN feedback begins 
to redistribute gas and suppress star formation \cite{Schaller:2014uwa,Mummery:2017lcn}. 
However, in Appendix~\ref{app:tngdark} we compare the inferred bias parameters from \texttt{TNG300} with those from its dark-matter-only counterpart, \texttt{TNG300-Dark}. The close agreement between the two indicates that baryonic effects in \texttt{TNG300} are unlikely to be the dominant origin of the HV--TNG offsets discussed here.

The smaller 
TNG300 volume ($L\simeq205\,h^{-1}$Mpc) means the massive-halo 
population is more sparsely sampled and subject to larger sample 
variance, introducing additional scatter in the field-level 
bias measurements from that smaller effective volume. The 
FastPM approximate gravity solver used in HV can also affect 
halo concentrations and internal profiles at the high-mass end 
relative to a full tree-code integration \cite{Feng:2016yqz,Modi:2019ewx}. These effects compound 
at large $\alpha$ because up-weighting the massive-halo 
contribution turns small fractional differences in the halo 
population into large differences in the effective 
\HI-weighted biases.

The practical implication for simulation-based priors is that the mapping from HOD parameters to EFT bias parameters retains a non-negligible dependence on the underlying simulation, especially in the high-bias tail and in the tidal sector. For broad informative priors, this level of difference may still be acceptable and would likely remain far more useful than adopting wide uniform priors with no simulation input. For precision applications, however, the HV--TNG differences are large enough that they should be treated as a real modeling systematic. In the next section, we explore the implications of these differences on Fisher forecasts involving the 21\,cm power spectrum.

\section{Propagating CHORD-like 21\,cm constraints to EFT priors}\label{sec:flow_usage}

We now show one concrete use of the conditional density learned above. Starting from a forecast for the 21\,cm power spectrum measurement at nonlinear scales, we infer a posterior on the HOD parameters and then propagate it through the conditional flow to obtain a prior on the EFT bias parameters.

At each redshift, we define the real-space 21\,cm power spectrum as
\begin{equation}
P_{21}(k,z) = \bar T_b^2(z)\,P_{\rm HI}(k,z)\ ,
\end{equation}
where $P_{\rm HI}(k,z)$ is the real-space power spectrum of the \HI\ overdensity field measured from the Hidden Valley realizations, and we use fitting functions from Refs.~\cite{CHIME:2022kvg} and~\cite{Crighton:2015pza} for the mean 21\,cm brightness temperature $\bar{T}_{\rm b}(z)$ and the \HI\ density parameter $\Omega_{\rm HI}(z)$:
\begin{align}
\bar{T}_{\rm b}(z) &=191.06\,\frac{h}{E(z)}\,\Omega_{\rm HI}^{\rm fid}(z)\,(1+z)^2\ {\rm mK}\ , \\
\Omega_{\rm HI}^{\rm fid}(z) &= 4\times10^{-4}(1+z)^{0.6}\ .
\end{align}
Thus, in this forecast exercise we keep the mean \HI\ abundance fixed through a fiducial $\Omega_{\rm HI}(z)$ and propagate only the HOD dependence of the clustering signal.
Because the calculation is performed in real space, it neglects redshift-space effects such as Finger-of-God damping. At $z\gtrsim 1.5$, the majority of CHORD's sensitivity in the wavenumber in our forecasts will come from modes with $k_\parallel \gg k_\perp$ (e.g.~\cite{Foreman:2024kzw}), for which Finger-of-God damping will be a strong effect; therefore, the signal-to-noise in these forecasts should be regarded as optimistic.

We choose a fiducial HOD point
\begin{align}
\nonumber
\thetaHOD^\star &= (M_0,M_{\rm min},\alpha,\beta) \\
&=(3\times10^{10},\,2\times10^{11},\,0.8,\,0.6)\ ,
\end{align}
and use the subset of forecast bins returned by the CHORD sensitivity calculation that lie in the range
\begin{equation}
0.5\,h\,{\rm Mpc}^{-1}<k<1.2\,h\,{\rm Mpc}^{-1}\ .
\end{equation}
We then approximate the HOD Fisher matrix as
\begin{equation}
F^{\rm(HOD)}_{ij}
=
\sum_{\mu}
\frac{1}{\sigma_P^2(k_\mu)}
\left.
\frac{\partial P_{21}(k_\mu)}{\partial \theta_{{\rm HOD},i}}
\right|_{\thetaHOD^\star}
\left.
\frac{\partial P_{21}(k_\mu)}{\partial \theta_{{\rm HOD},j}}
\right|_{\thetaHOD^\star},
\label{eq:Fisher_HOD_P21}
\end{equation}
where $\mu$ runs over the discrete logarithmic $k$ bins with width
$\Delta\log k\simeq 0.308$. 
Here $\sigma_P(k_\mu)$ is taken from a CHORD sensitivity forecast produced with \texttt{py21cmsense}\footnote{\href{https://github.com/rasg-affiliates/21cmSense.git}{https://github.com/rasg-affiliates/21cmSense.git}}~\cite{Pober:2013jna,Liu:2019awk}, assuming the code's \texttt{moderate} foreground model. In this model, modes below the horizon wedge plus a fixed buffer are removed. Writing $k_\parallel$ for the magnitude of the line-of-sight wavenumber, the excised modes satisfy
\begin{equation}
k_\parallel < \beta_{\rm hor}(z)\,k_\perp + B ,
\end{equation}
where
\begin{equation}
\beta_{\rm hor}(z)
=
\frac{\chi(z)H(z)}{c(1+z)}
\end{equation}
is the horizon-wedge slope, and
\begin{equation}
B = 0.1\,h\,{\rm Mpc}^{-1}
\end{equation}
is the additive buffer. Here $\chi(z)$ is the transverse comoving distance, $H(z)$ is the Hubble rate, and $c$ is the speed of light. The instrumental parameters used in the sensitivity calculation are given in Appendix~\ref{app:chord}.

In Eq.~\eqref{eq:Fisher_HOD_P21} we make two simplifying assumptions: we take the covariance to be diagonal in $k$, and we use the instrumental uncertainty returned by the sensitivity code as the variance.
This ignores sample variance and off-diagonal mode coupling, and is another reason why these forecasts should be viewed as optimistic, especially at the lowest redshift where the signal is strongest.

The derivatives in Eq.~\eqref{eq:Fisher_HOD_P21} are evaluated numerically by central differences using nine Hidden Valley runs constructed around $\thetaHOD^\star$, with steps
\begin{equation}
\Delta M_0/M_0=\Delta M_{\rm min}/M_{\rm min}=0.05\ ,
\quad
\Delta\alpha=\Delta\beta=0.05\ .
\end{equation}
In practice, the Fisher matrix is poorly conditioned unless we regularize the weakest directions. We therefore impose broad Gaussian priors on $M_0$ and $M_{\rm min}$, with widths equal to 30\% and 10\% of their training ranges, respectively. These priors are intended to represent external information that could, in principle, be supplied by independent observations: $M_0$ is closely related to the overall \HI\ abundance, while $M_{\rm min}$ controls the characteristic halo-mass scale at which halos begin to host significant \HI. In the present proof-of-concept forecast, we use broad priors simply to regularize directions that are otherwise weakly constrained or partially degenerate in the Fisher analysis, especially the overall normalization direction associated with $M_0$. We then take
\begin{equation}
{\sf C}_{\rm HOD}=(F^{\rm(HOD)})^{-1}
\end{equation}
as an approximate covariance for the HOD posterior.

We approximate the HOD posterior as
\begin{equation}
p(\thetaHOD)\simeq \mathcal{N}(\thetaHOD^\star,{\sf C}_{\rm HOD})\ ,
\end{equation}
draw $N=2000$ HOD samples from this Gaussian, and clip them to the training range of the Hidden Valley scan to avoid extrapolating the flow.
For each draw $\thetaHOD^{(n)}$, we sample
\begin{equation}
\thetaEFT^{(n)}\sim p(\thetaEFT\mid\thetaHOD^{(n)})
\end{equation}
using the conditional normalizing flow, with 25 EFT samples per HOD draw.
This gives $5\times10^4$ samples of
$(b_1,b_2,b_{\mathcal G_2},b_3)$
for each observing scenario and redshift, which we use to visualize the induced EFT prior.

\begin{figure*}
    \centering
    \includegraphics[width=\linewidth]{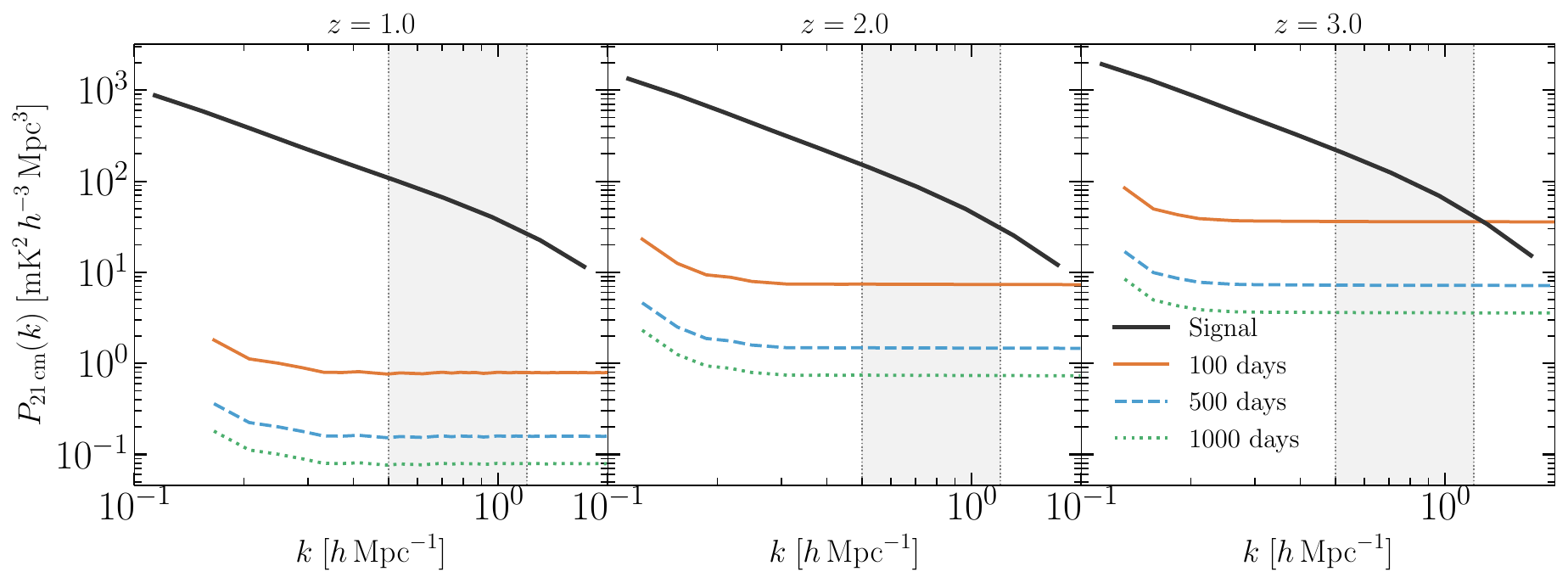}
    \caption{Real-space 21\,cm signal power spectrum and CHORD noise levels used in the Fisher forecast at $z=1$, $2$, and $3$.
    The black curve shows the fiducial signal $P_{21}(k)=\bar T_b^2 P_{\rm HI}(k)$, and the colored curves show the forecast power-spectrum uncertainty for 100, 500, and 1000 observing days using the \texttt{moderate} foreground model of \texttt{py21cmsense}.
    The $k$ range used in the Fisher calculation, $0.5<k/(h\,{\rm Mpc}^{-1})<1.2$ (shaded in grey), lies in a high signal-to-noise regime at $z=1$, an intermediate regime at $z=2$, and a substantially noisier regime at $z=3$.
    This progression explains why 
    the EFT priors induced by HOD constraints on these scales
    are already tight at $z=1$ but remain broad at $z=3$ for short observing times.}
    \label{fig:chord_signal_noise}
\end{figure*}

Figure~\ref{fig:chord_signal_noise} gives the basic reason for the redshift dependence of the final EFT priors. At $z=1$, the fiducial signal lies comfortably above the forecast uncertainty over most of the $k$ range used in the Fisher matrix. At $z=2$, the separation is smaller but still favorable. At $z=3$, the 100-day case becomes much more noise-limited, especially toward the upper end of the range.

\begin{figure}
    \centering
    \includegraphics[width=\linewidth]{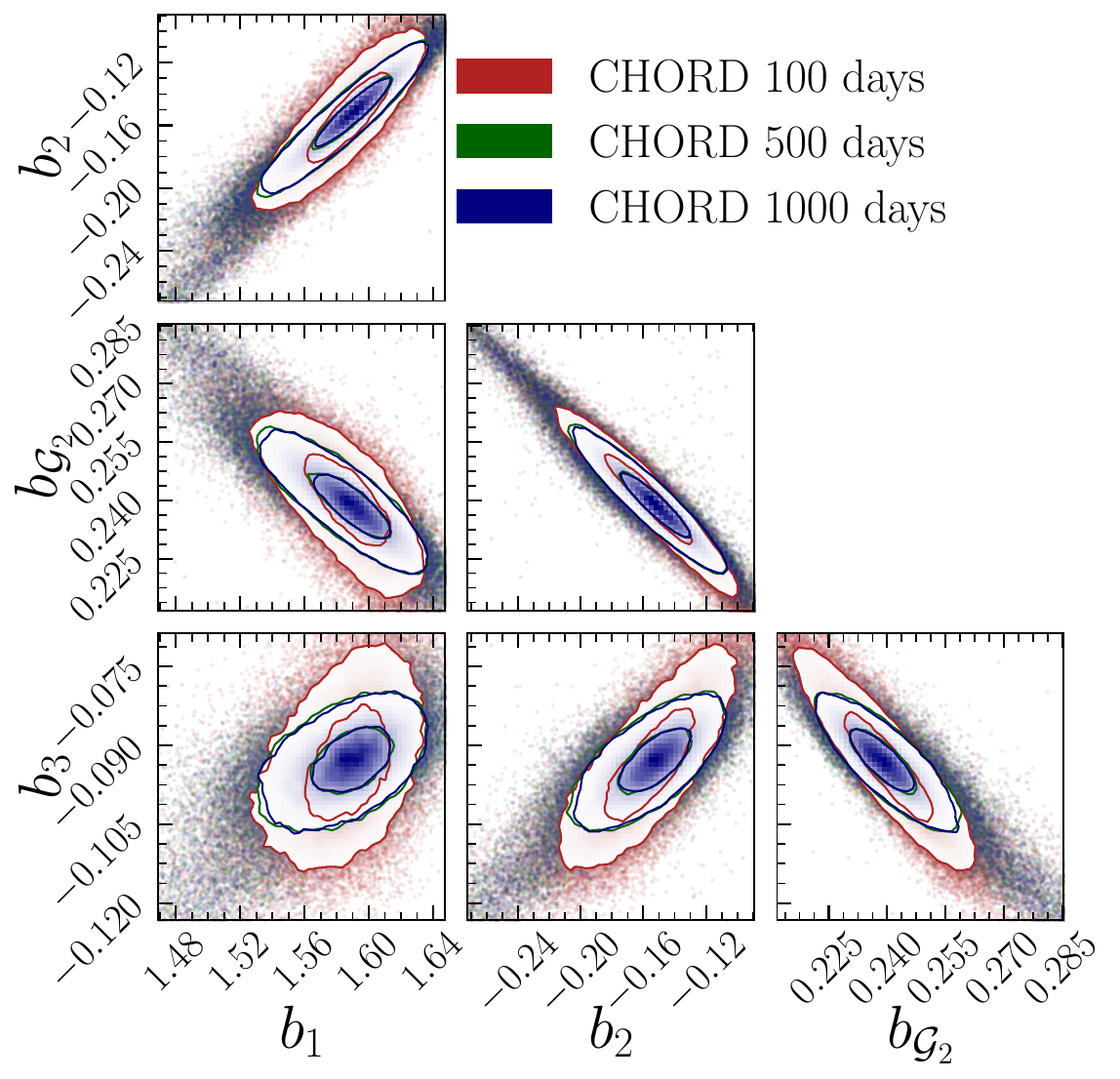}
    \caption{Simulation-based EFT priors induced by CHORD-like 21\,cm power-spectrum constraints at $z=1$.
    Red, green, and navy contours correspond to 100, 500, and 1000 days of observation, enclosing 68\% and 95\% of the probability mass.
    The fiducial HOD is $(M_0,M_{\rm min},\alpha,\beta)=(3\times10^{10}\,h^{-1}\,M_\odot,2\times10^{11}\,h^{-1}\,M_\odot,0.8,0.6)$.
    At this redshift the three scenarios already give compact, nearly elliptical priors, with only modest tightening between 100 and 1000 days.}
    \label{fig:chord_bias_corner_z1}
\end{figure}

\begin{figure}
    \centering
    \includegraphics[width=\linewidth]{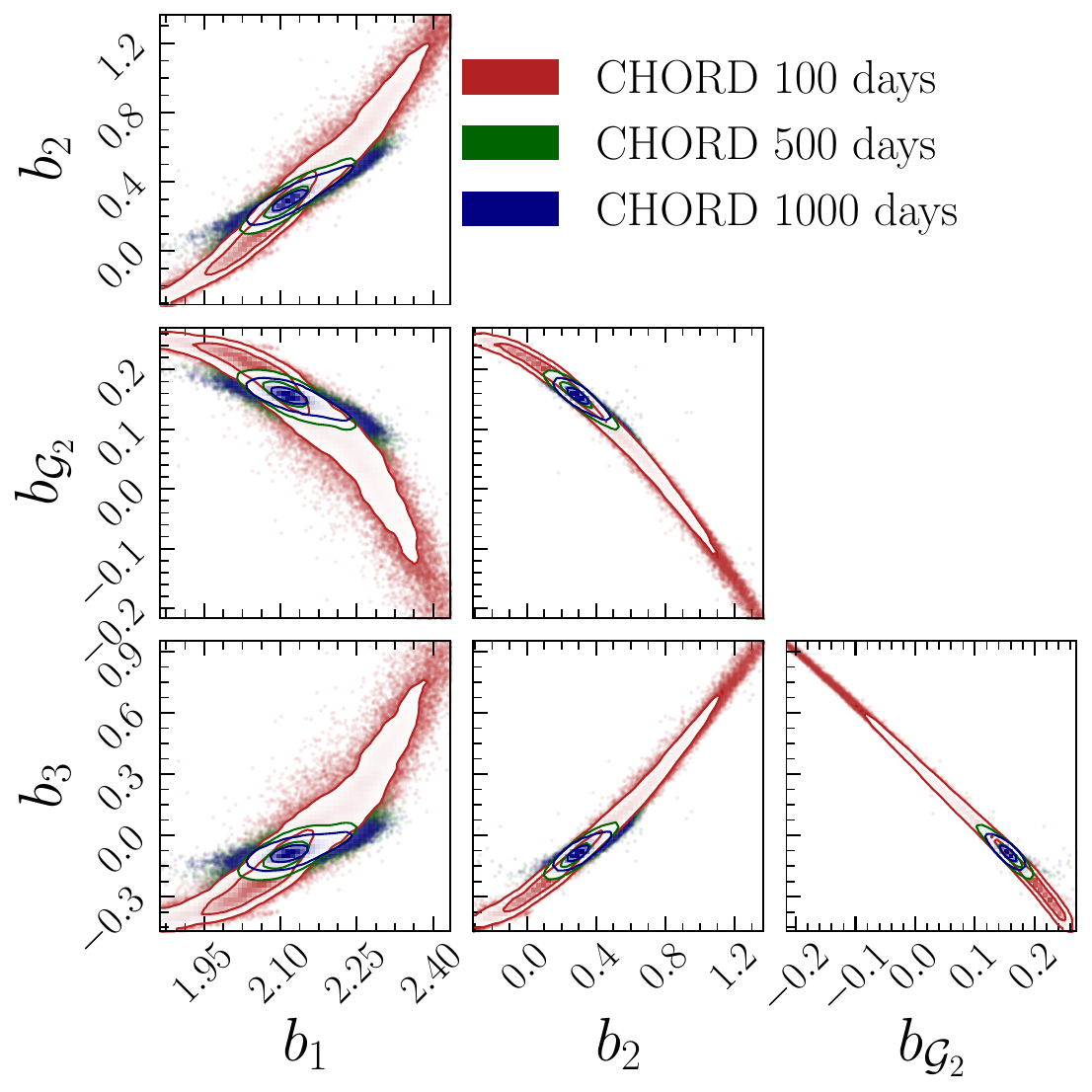}
    \caption{Same as Fig.~\ref{fig:chord_bias_corner_z1}, but at $z=2$.
    The priors remain fairly compact, but the separation between the 100-, 500-, and 1000-day scenarios is more visible than at $z=1$, showing that the gain from additional observing time is becoming more important.}
    \label{fig:chord_bias_corner_z2}
\end{figure}

\begin{figure}
    \centering
    \includegraphics[width=\linewidth]{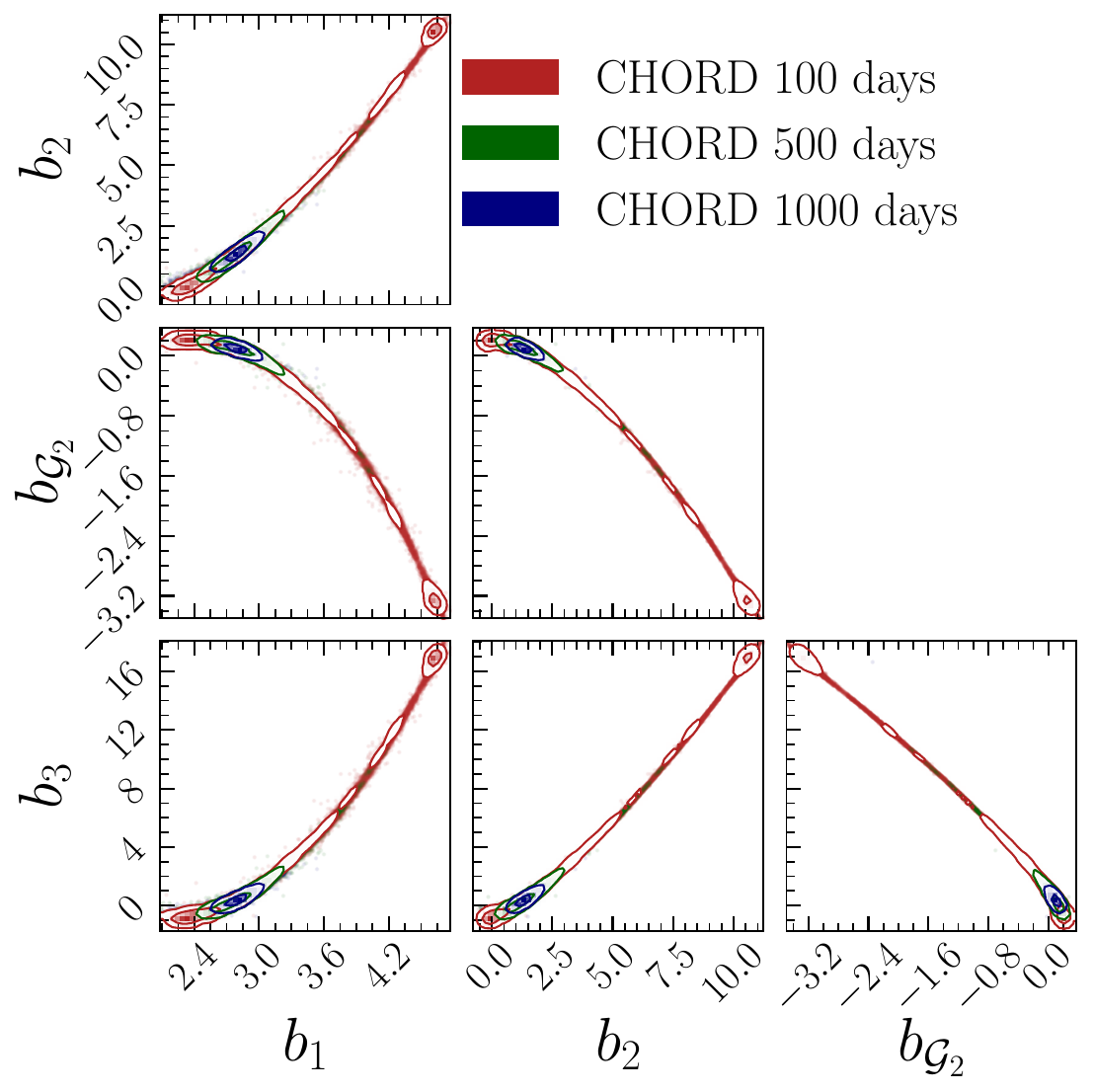}
    \caption{Same as Fig.~\ref{fig:chord_bias_corner_z1}, but at $z=3$.
    This is close to the high-redshift end of the range accessible to CHORD and SKA-Mid, and it is the most demanding of the three cases shown here.
    The 100-day contours are broad and visibly non-Gaussian, while 500 and 1000 days tighten the prior substantially.
    }
    \label{fig:chord_bias_corner_z3}
\end{figure}

We take the $z=1$ result in Fig.~\ref{fig:chord_bias_corner_z1} as the primary example, since this is the regime most closely connected to current post-reionization 21\,cm measurements. Even 100 days already localizes the EFT parameters to a compact region, and increasing the observing time mainly sharpens that region rather than changing its orientation. At this redshift, a Gaussian approximation to the induced prior would be fairly reasonable by eye, although the flow still retains the exact shape and correlations of the simulation-based mapping.

At $z=2$ (Fig.~\ref{fig:chord_bias_corner_z2}), the same general picture holds, but the dependence on observing time is clearer. The 500- and 1000-day contours are visibly tighter than the 100-day contours, indicating that the HOD posterior is no longer saturated by the chosen $k$ range. The induced EFT prior is still fairly regular in shape, but it is broader and more anisotropic than at $z=1$.

The strongest non-Gaussianity appears at $z=3$ (Fig.~\ref{fig:chord_bias_corner_z3}), which is also one of the highest redshifts relevant for CHORD and SKA-Mid. Here the 100-day prior spreads over a much larger fraction of the allowed bias manifold, and its shape is clearly not well described by a single Gaussian. Going to 500 and 1000 days tightens the prior substantially, especially in the $b_2$ and $b_3$ directions, but residual curvature remains visible. This is the regime in which the flow description is most useful.

\begin{figure*}
\centering
\includegraphics[width=\linewidth]{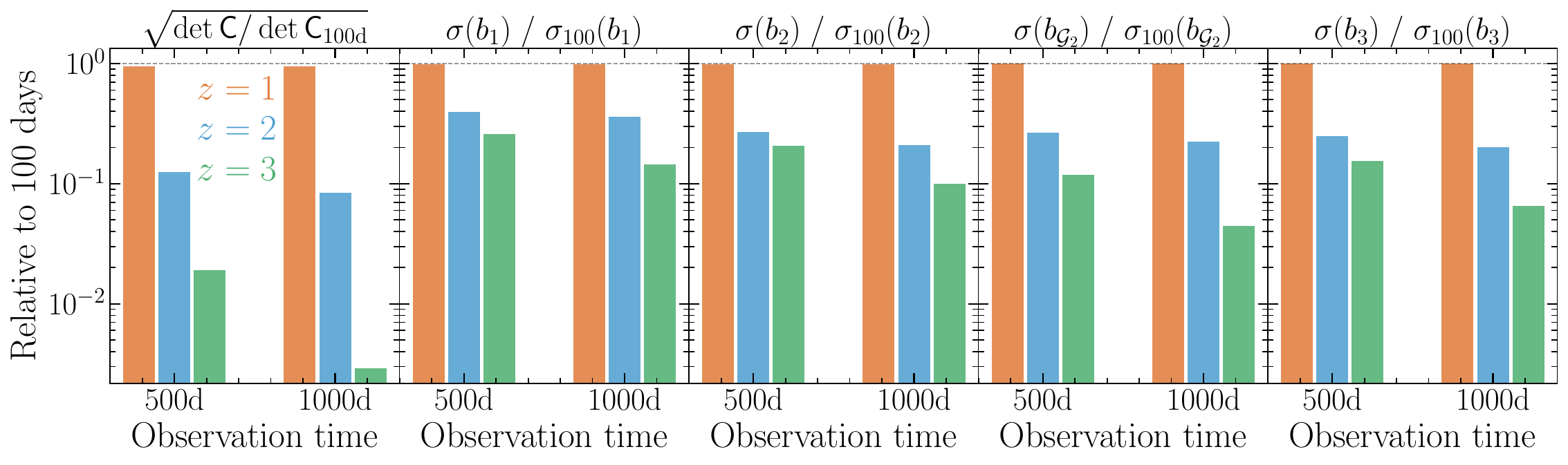}
\caption{Relative tightening of the induced EFT priors with CHORD observing time at $z=1$, $2$, and $3$, normalized to the corresponding 100-day result at each redshift. The first panel shows the Gaussian-equivalent covariance-volume ratio $\sqrt{\det{\sf C}/\det{\sf C}_{100{\rm d}}}$, where ${\sf C}$ is the sample covariance matrix of the propagated EFT samples. The remaining panels show the marginalized error ratios for $b_1$, $b_2$, $b_{\mathcal G_2}$, and $b_3$. At $z=1$, the ratios remain close to unity, indicating that the induced prior is already close to saturated after 100 days. At $z=2$, increasing the observing time produces a clear reduction in both the total covariance volume and the marginalized errors. The strongest improvement occurs at $z=3$, where the 100-day prior is broad and additional observing time substantially reduces the allowed EFT-prior volume.}
\label{fig:volume_reduction}
\end{figure*}

To summarize the gain from additional observing time more compactly, we compute the sample covariance matrix ${\sf C}$ of the propagated EFT samples for each redshift and observing scenario. Figure~\ref{fig:volume_reduction} shows the Gaussian-equivalent covariance-volume ratio,
\begin{equation}
\frac{V}{V_{100{\rm d}}}
=
\left(\frac{\det {\sf C}}{\det {\sf C}_{100{\rm d}}}\right)^{1/2},
\end{equation}
together with the marginalized error ratios
$\sigma(b_i)/\sigma_{100{\rm d}}(b_i)$ for
$i=1,2,\mathcal{G}_2,3$. All quantities are normalized to the corresponding 100-day result at the same redshift, so values below unity indicate tightening relative to the 100-day case.

Figure~\ref{fig:volume_reduction} quantifies the visual trends seen in Figs.~\ref{fig:chord_bias_corner_z1}--\ref{fig:chord_bias_corner_z3}. At $z=1$, the volume and marginalized-error ratios stay close to one, consistent with the nearly overlapping contours in Fig.~\ref{fig:chord_bias_corner_z1}. This indicates that, within the assumptions of this forecast, the induced EFT prior is already largely saturated after 100 days. At $z=2$, the 500- and 1000-day cases show a more visible reduction in both the covariance volume and the individual bias uncertainties. The effect is strongest at $z=3$: the covariance volume decreases by more than an order of magnitude by 500 days and by roughly two orders of magnitude by 1000 days, while all four marginalized bias errors decrease coherently. This behavior follows the signal-to-noise hierarchy in Fig.~\ref{fig:chord_signal_noise}: the lower-redshift cases are already well constrained with 100 days, whereas the high-redshift case remains noise limited and therefore benefits most from additional observing time.

\begin{figure}
\centering
\includegraphics[width=\linewidth]{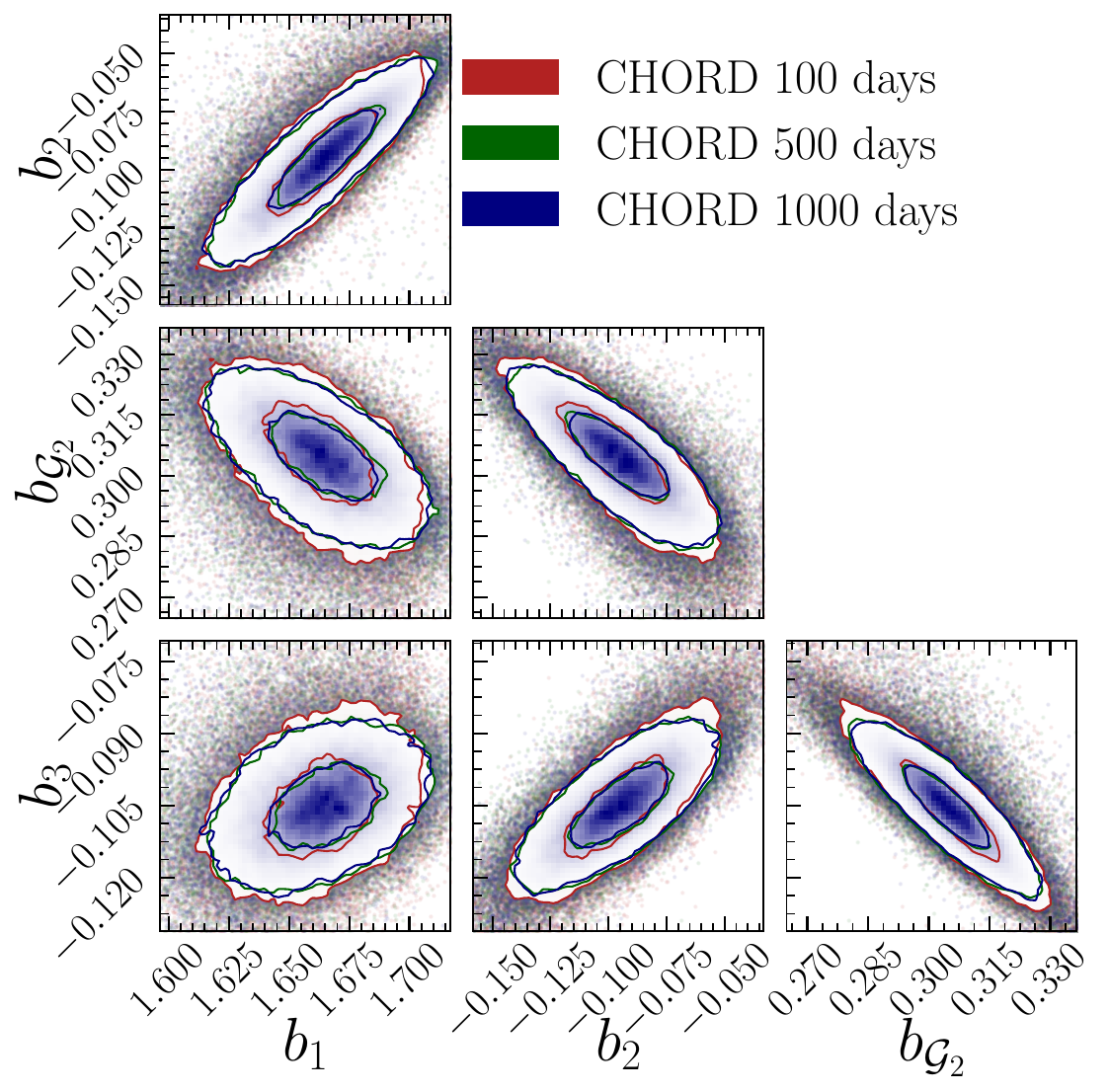}
\caption{IllustrisTNG300 analogue of Fig.~\ref{fig:chord_bias_corner_z1}: simulation-based EFT priors induced by the same CHORD-like $z=1$ 21,cm power-spectrum constraints, but propagated through the conditional flow trained on \texttt{TNG300} rather than Hidden Valley. Red, green, and navy contours correspond to 100, 500, and 1000 observing days, enclosing 68\% and 95\% of the probability mass. The contours remain compact and nearly elliptical, as in the Hidden Valley case, but their location in EFT-parameter space is systematically shifted.}
\label{fig:tng_chord_bias_corner_z1}
\end{figure}

As a check on the simulation dependence of this propagation step, we repeat the same $z=1$ exercise using the conditional flow trained on the IllustrisTNG300 HOD--bias pairs. The result is shown in Fig.~\ref{fig:tng_chord_bias_corner_z1}. The forecast setup is kept the same as in Fig.~\ref{fig:chord_bias_corner_z1}; the only change is that the HOD posterior is pushed through the TNG300-trained conditional density rather than the Hidden-Valley-trained one. The comparison, therefore, isolates the effect of changing the learned HOD-to-bias mapping.

The TNG300-induced prior has a broadly similar shape to the Hidden Valley prior, but it is centered in a noticeably different region of EFT-parameter space. From the plotted contours, the \texttt{TNG300} prior is centered approximately around
$(b_1,b_2,b_{\mathcal G_2},b_3)
\simeq
(1.65,,-0.10,,0.30,,-0.10),$
whereas the corresponding Hidden Valley result in Fig.~\ref{fig:chord_bias_corner_z1} is centered roughly around
$(b_1,b_2,b_{\mathcal G_2},b_3)
\simeq
(1.58,,-0.16,,0.245,,-0.093).$
Despite these offsets, the covariance structure is similar in the two cases. The signs of the correlations are also preserved. The main effect of replacing the Hidden-Valley-trained flow with the TNG300-trained flow is therefore not a dramatic change in the shape of the prior, but a coherent translation of the prior support in EFT-parameter space.

This comparison reinforces the conclusion of Sec.~\ref{sec:tng}: the qualitative structure of the HOD-to-bias mapping is robust, but the quantitative mapping is not completely simulation independent. For broad informative priors, either simulation-based prior would still be much more restrictive than an uninformative flat prior. For precision applications, however, the HV--TNG displacement should be treated as a modeling systematic.

Taken together, Figs.~\ref{fig:chord_signal_noise}--\ref{fig:volume_reduction} provide a proof-of-concept implementation of one possible use of the learned HOD-to-bias mapping. In this mock analysis, nonlinear-scale \tcm power-spectrum information is used to construct an approximate posterior on the HOD parameters, and this posterior is then pushed through the conditional flow to obtain an induced prior on the EFT bias parameters. More generally, the same flow can be used in two ways: directly as a simulation-based prior in quasi-linear EFT analyses, or as a map from external or nonlinear-scale constraints on \HI\ astrophysics to EFT-parameter priors. The exercise shown here follows the latter route. The resulting priors therefore inherit the assumptions of the present setup, including the simplified HOD model, real-space power spectra, fixed $\Omega_{\rm HI}(z)$, and diagonal forecast covariance. Within these caveats, the calculation demonstrates that nonlinear-scale information about \HI\ clustering can be translated into informative, correlated, priors on the EFT bias parameters.

\section{Summary and discussion}\label{sec:summary}

We have presented a proof-of-concept framework for constructing simulation-based priors on the EFT bias parameters of post-reionization \HI\ intensity maps. The central object is the conditional distribution
$p(\thetaEFT\mid\thetaHOD)$, where
$\thetaHOD=(M_0,M_{\rm min},\alpha,\beta)$
specifies a simplified \HI--HOD model and
$\thetaEFT=(b_1,b_2,b_{\mathcal G_2},b_3)$
denotes the large-scale bias parameters inferred from field-level transfer-function fits. This connects two ideas that have so far developed mostly separately: field-level measurements of \HI\ bias parameters using shifted-operator methods~\cite{Schmittfull:2018yuk,Schmittfull:2020trd,Obuljen:2022cjo,Foreman:2024kzw}, and the use of simulation-based priors in EFT analyses of biased tracers~\cite{Ivanov:2024hgq,Ivanov:2024xgb,DESI:2025wzd,Chen:2025jnr,Chudaykin:2026nls}.

The framework has two natural use cases. First, the learned distribution can be used directly as a simulation-based prior in quasi-linear EFT analyses of \HI\ clustering, in analogy with recent galaxy-survey applications. Second, external or nonlinear-scale measurements that constrain $\thetaHOD$\ can be propagated through the flow to obtain an induced prior on $\thetaEFT$. The CHORD-like calculation in Sec.~\ref{sec:flow_usage} is a mock implementation of the second route: a simplified nonlinear-scale \tcm power-spectrum forecast is used to build an approximate HOD posterior, which is then pushed through the trained flow.

Our first main result is that the \HI\ HOD-to-bias mapping is highly structured. Across the Hidden Valley ensemble, the inferred bias parameters occupy a narrow, curved, and non-Gaussian region of $\thetaEFT$\ space rather than filling the four-dimensional prior volume independently. The high-mass slope $\alpha$ is strongly correlated with position along this bias manifold: increasing $\alpha$ generally moves the \HI\ weight toward more massive halos, increasing $b_1$, $b_2$, and $b_3$, while making $b_{\mathcal G_2}$ more negative. The parameters $M_{\rm min}$ and $\beta$ have weaker but still visible effects, mostly changing the thickness and detailed shape of the bias--bias relations, while $M_0$ has little direct impact on the deterministic overdensity field because an overall \HI\ mass normalization largely cancels in $\delta_{\rm HI}$. The redshift evolution shown in Appendix~\ref{app:b2_allz} indicates that this curved bias--bias structure persists across the post-reionization redshift range studied here.

We also compared the field-level measurements with simple \HI-mass-weighted halo-bias expectations. These analytic estimates capture part of the broad ordering of the simulation measurements, but they do not reproduce the detailed relations. The discrepancy is visible already for $b_1$ and $b_2$, and is especially pronounced for the effective cubic response $b_3$. This is not surprising, since the measured $b_3$ is the large-scale coefficient of the cubic local operator within our truncated field-level basis and can absorb contributions from omitted third-order operators. More broadly, this comparison shows that mass-weighted halo-bias formulae are useful baselines, but are not sufficient to define accurate EFT priors across the full HOD domain.

We trained conditional normalizing flows to model $p(\thetaEFT\mid\thetaHOD)$ directly. The single-parameter flow sweeps show that the trained density estimator captures both the mean HOD response and the HOD-dependent width of the conditional distribution. The resulting model is therefore sampleable, correlated, and non-Gaussian by construction. This is important for downstream EFT analyses: replacing broad, independent priors with a flow-based prior preserves the curved support of the simulation ensemble, rather than approximating it with independent Gaussian widths or fixed one-parameter relations in $b_1$.

We tested the dependence of the mapping on the underlying simulation by repeating the analysis with IllustrisTNG300. The qualitative structure is similar to Hidden Valley: the bias parameters again lie on curved manifolds, and the hierarchy of HOD sensitivities is broadly preserved. Quantitatively, however, the two simulations do not give identical mappings. The largest differences appear in the high-bias tail and in the tidal sector. Mass-restricted comparisons show that the high-mass halo population strongly affects the HV--TNG offsets, although removing the high-mass tail does not eliminate them completely. The comparison with \texttt{TNG300-Dark} in Appendix~\ref{app:tngdark} further suggests that baryonic effects on halos in \texttt{TNG300} are not the dominant origin of these offsets. Other simulation-level differences, including volume, halo abundance, mass calibration, resolution, and gravity solver, are therefore likely to contribute. For precision applications, this simulation dependence should be treated as a modeling systematic.

The CHORD-like mock analysis illustrates how nonlinear-scale information could be translated into an EFT prior using the learned conditional density. At $z=1$, the induced prior is already compact after 100 observing days, and increasing the observing time mostly leaves the contour shape unchanged. At $z=2$, the improvement from 100 to 500--1000 days is more visible. At $z=3$, where the forecast signal-to-noise is lower, the 100-day induced prior is broad and visibly non-Gaussian, while longer observing times substantially reduce both the covariance volume and the marginalized errors. Repeating the same exercise with the IllustrisTNG300-trained flow at $z=1$ gives contours with similar covariance structure but shifted mean location, consistent with the simulation-dependence seen in Sec.~\ref{sec:tng}. These results demonstrate the potential value of \HI-specific simulation-based priors, while also showing that the final prior can depend on the training simulation.

The present analysis is designed as a controlled first implementation, and several refinements would make the resulting priors more directly applicable to data. First, the HOD model in Eq.~\eqref{eq:HIHOD} is intentionally simple: it depends only on halo mass and ignores scatter at fixed mass, assembly bias, separate central and satellite contributions, and the spatial distribution of \HI\ within halos. These effects are known to be relevant for simulation-based priors in galaxy surveys~\cite{Akitsu:2024lyt,Shiferaw:2024ehr}, and they may also affect \HI\ priors. They can be incorporated in the present framework by enlarging $\thetaHOD$\ and regenerating the training ensemble. Second, all field-level bias measurements here are performed in real space. Observed \HI\ maps are affected by redshift-space distortions \cite{Sarkar:2018gcb, Sarkar:2019nak}, Finger-of-God damping \cite{Jackson:1971sky}, foreground cuts, and additional EFT nuisance parameters; extending the shifted-operator pipeline to redshift space is therefore a necessary next step. Third, the CHORD-like calculation uses a diagonal forecast covariance, fixed $\Omega_{\rm HI}(z)$, and a Gaussian Fisher posterior for $\thetaHOD$. These assumptions make the calculation a controlled demonstration, but the resulting priors should not be interpreted as final survey-ready priors.

The simulation comparison points to another important improvement: future training sets should include multiple simulations with controlled differences in volume, resolution, gravity solver, cosmology, and baryonic physics. Ideally, matched hydrodynamical and dark-matter-only simulations would allow the separate effects of baryons and numerical choices to be isolated. The present \texttt{TNG300}--\texttt{TNG300-Dark} comparison suggests that baryons are not the leading explanation for the offsets seen here, but a broader suite would be needed to turn that statement into a robust uncertainty model. Such simulation variation could be incorporated either by broadening the prior itself or by adding simulation-level hyperparameters to the conditional density.

There are also several direct extensions of the method. A more realistic analysis would replace the Fisher approximation with samples from an actual nonlinear-scale likelihood for $\thetaHOD$, then propagate those samples through the same flow. The current fixed-cosmology setup could be generalized to learn
$p(\thetaEFT\mid\thetaHOD,\theta_{\rm cosmo})$,
allowing cosmology and \HI\ astrophysics to vary jointly. The flow prior should also be inserted into an EFT likelihood to quantify the impact on cosmological constraints relative to broad uncorrelated priors. Finally, bispectrum and higher-order analyses \cite{Sarkar:2019ojl,Mazumdar:2020bkm,Pal:2026hkq} are a natural application, because they involve a larger bias-parameter space and therefore stand to benefit especially from informative simulation-based priors; this is particularly relevant for \HI\ forecasts of primordial non-Gaussianity~\cite{Karagiannis:2019jjx, Bharadwaj:2020wkc}.

In summary, the HOD-to-bias relation for \HI\ is nonlinear, correlated, and sensitive to how neutral hydrogen weights the halo population. Conditional normalizing flows provide an efficient way to encode this relation without reducing it to a Gaussian approximation or a fixed bias--$b_1$ relation. While the present analysis is deliberately simplified, it demonstrates that simulation-based information about the \HI--halo connection can be translated into informative priors on the EFT parameters that enter quasi-linear analyses of \HI\ intensity maps.

\acknowledgments

We thank Andrej Obuljen for initial collaboration on this project, and for assistance with the \texttt{Hi-Fi Mocks} code. 
We are deeply grateful to Adrian C.\ Liu for his sustained guidance throughout this project, including many insightful discussions, careful readings of the manuscript, detailed comments, and valuable suggestions that substantially shaped the analysis and improved the presentation of the work.
We thank Chirag Modi, Emanuele Castorina, Yu Feng, Martin White for making the Hidden Valley simulations publicly available, and the TNG Collaboration for making the IllustrisTNG simulations available.
This material is based upon work supported by the U.S.\ Department of Energy, Office of Science, Office of High Energy Physics under Award Number DE-SC0024309.
DS acknowledges the support of the Canada 150 Chairs program, the Fonds de recherche du Québec Nature et Technologies (FRQNT) and the Natural Sciences and Engineering Research Council of Canada (NSERC) joint NOVA grant, and the Trottier Space Institute Postdoctoral Fellowship program.
This research was enabled in part by computational resources and support provided by Calcul Qu\'ebec (\url{https://www.calculquebec.ca}) and the Digital Research Alliance of Canada (\url{https://alliancecan.ca}).

\appendix

\section{Redshift evolution of bias--$b_1$ relations}\label{app:b2_allz}

\begin{figure*}
    \centering
    \includegraphics[width=0.7\textwidth]{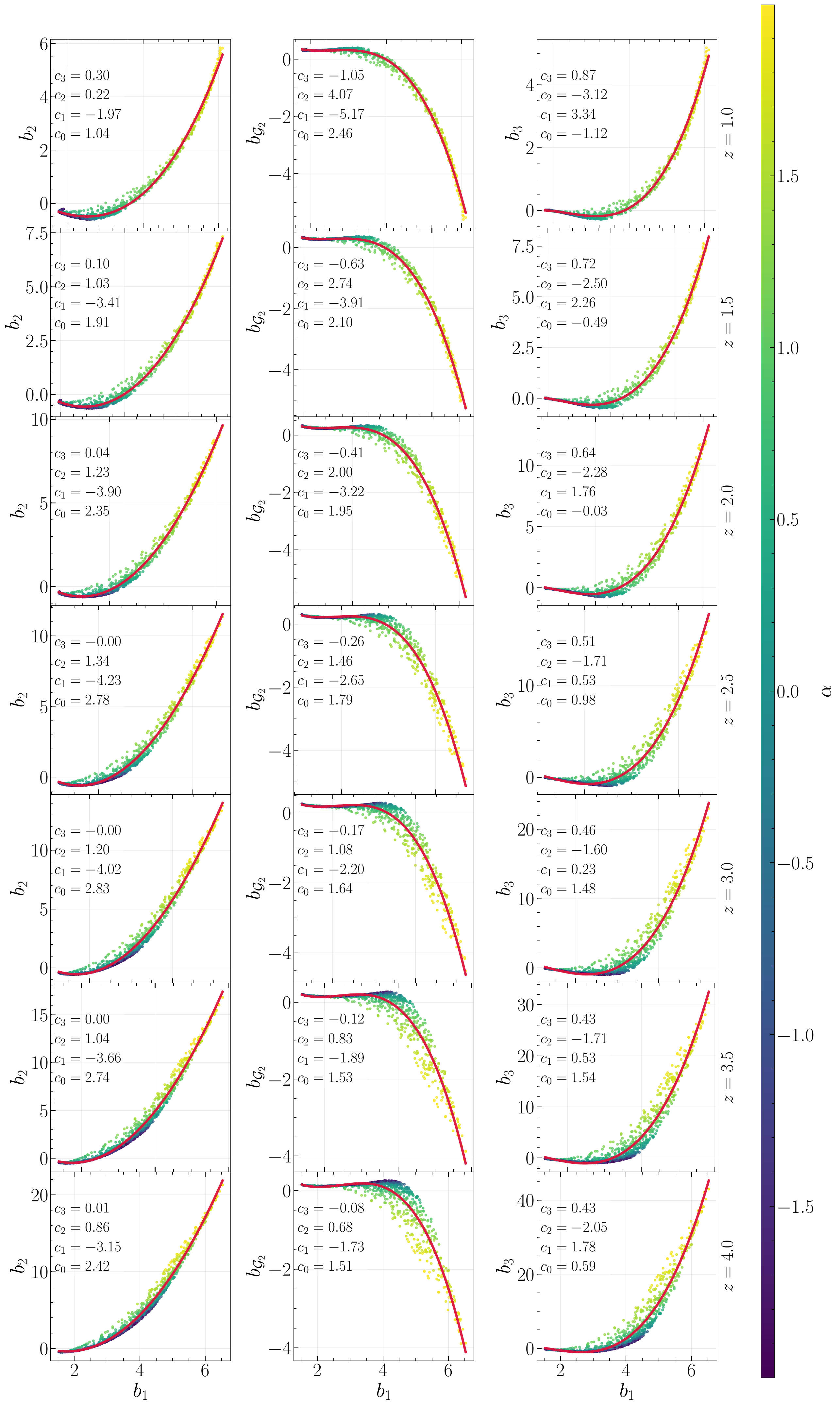}
    \caption{Hidden Valley: $b_2$ (left), $b_{\mathcal G_2}$ 
    (middle), and $b_3$ (right) as functions of $b_1$ across all 
    seven available redshifts (rows, $z=1.0$ to $4.0$ top to 
    bottom).
    Points are color-coded by $\alpha$; red curves show the 
    best-fit cubic polynomial (Eq.~\ref{eq:bias_poly}) with 
    coefficients printed in each panel.
    The dynamic range of all three biases grows monotonically 
    with redshift, and the $\alpha$-dependent scatter around the 
    mean relations persists across the full redshift range.
    }
    \label{fig:bias_b1_allz_hv}
\end{figure*}

\begin{figure*}
    \centering
    \includegraphics[width=0.7\textwidth]{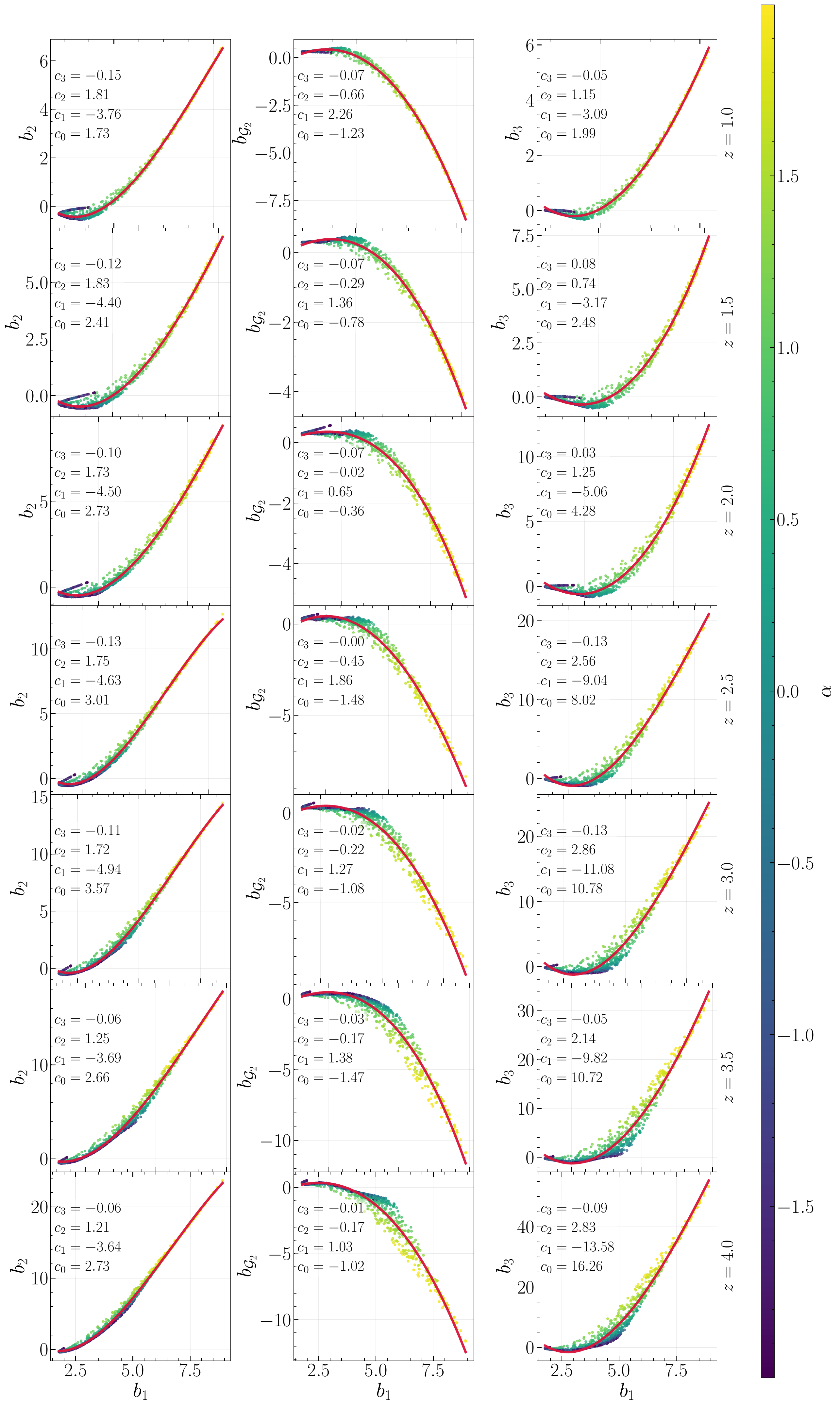}
    \caption{IllustrisTNG300: same layout as 
    Fig.~\ref{fig:bias_b1_allz_hv}.
    The $b_2(b_1)$ and $b_3(b_1)$ relations follow qualitatively 
    similar nonlinear loci to Hidden Valley, but \texttt{TNG300} extends 
    to larger $b_1$ ($\lesssim10$) and shows markedly different 
    behavior in $b_{\mathcal G_2}(b_1)$: at most redshifts, 
    $b_{\mathcal G_2}$ is near zero or positive at intermediate 
    $b_1$ before turning negative at large $b_1$, in contrast to 
    the predominantly negative HV behavior. 
    }
    \label{fig:bias_b1_allz_tng}
\end{figure*}

Figures~\ref{fig:bias_b1_allz_hv} and~\ref{fig:bias_b1_allz_tng} 
extend the bias--$b_1$ relations of Sec.~\ref{sec:hv_hod_bias} 
across all seven redshifts available in the Hidden Valley and 
IllustrisTNG300 ensembles ($z=1.0$, $1.5$, $2.0$, $2.5$, $3.0$, 
$3.5$, and $4.0$), showing $b_2(b_1)$, $b_{\mathcal G_2}(b_1)$, 
and $b_3(b_1)$ in the three columns respectively.
Points are color-coded by $\alpha$, and each panel overlays a 
cubic polynomial fit
\begin{equation}
b_a(b_1) = c_3\,b_1^3 + c_2\,b_1^2 + c_1\,b_1 + c_0\ ,
\label{eq:bias_poly}
\end{equation}
with the best-fit coefficients $(c_0,c_1,c_2,c_3)$ quoted in each 
panel. We provide this tabulation as a practical reference for 
analyses requiring a quick analytic approximation to the \HI\ 
bias relations across the redshift range relevant for 
post-reionization intensity mapping. 
However, we remind the reader of the limitations of our assumptions (ignoring redshift-space distortions, \HI\ assembly bias, the distinction between centrals and satellites, and the extended nature of \HI\ in halos), and caution that the quoted relations may change in a more realistic setup.


\section{CHORD Instrumental Parameters}
\label{app:chord}

The \HI\ power-spectrum noise model $\sigma_P(k)$ used in the Fisher
forecasts of Sec.~\ref{sec:flow_usage} is computed with
\texttt{py21cmSense}~\cite{Pober:2013jna,Liu:2019awk}, which models
the thermal noise power spectrum from the array configuration,
observing strategy, system temperature, beam model, and frequency resolution.
For a given $uv$ cell, the thermal-noise level is set by the accumulated integration time from all baselines that sample that cell. Schematically, the RMS noise scales as
\begin{equation}
\sigma_{\rm rms}(u,v)
\propto
\frac{T_{\rm sys}}{\sqrt{N_{\rm red}\,t_{\rm obs}\,\Delta\nu}}\,
\Omega_{\rm eff}\ \,
\end{equation}
where $T_{\rm sys}$ is the system temperature, $N_{\rm red}$ is the number of redundant baselines contributing to the same $uv$ cell, $t_{\rm obs}$ is the observing time assigned to that cell, $\Delta\nu$ is the channel width, and $\Omega_{\rm eff}$ denotes the appropriate beam-area factor for the adopted primary-beam model. The redundancy factor $N_{\rm red}$ counts baselines, not polarizations; the improvement from combining the two observed polarizations is included separately in the thermal-noise normalization used for Stokes-$I$ sensitivity.

In this calculation, the quoted observing time is assigned to a single CHORD-like drift track rather than being split among multiple independent declination strips. Splitting the same total observing time over several strips would reduce the integration time per strip and therefore increase the thermal-noise contribution for each strip. After the baseline sensitivities are computed, \texttt{py21cmSense} converts the surviving cells to $(k_\perp,k_\parallel)$, applies the foreground mask, and combines the remaining modes to obtain the 1D uncertainty $\sigma_P(k_\alpha)$ used in the Fisher matrix of Sec.~\ref{sec:flow_usage}. We use the \texttt{moderate} foreground model, which removes modes inside the horizon wedge with an additional $0.1\,h\,{\rm Mpc}^{-1}$ buffer in $k_\parallel$.

The CHORD-like instrumental parameters adopted in this calculation
are summarized in Table~\ref{tab:chord_params}. The forecast is evaluated separately at each redshift using $\nu=\nu_{21}/(1+z)$, so we do not quote a single central frequency.

\begin{table}[h]
\centering
\begin{tabular}{lc} 
\hline\hline 
\textbf{Parameter/Assumption} & \textbf{CHORD} \\ 
Dish diameter $D_{\rm dish}$ (m) & 6 \\ 
Antenna layout & Rectangular \\ 
Number of antennas & 512 ($32\times16$) \\ 
Integration time per visibility $t_{\rm int}$ (s) & 60 \\ 
Observing time per day (hr) & 8 \\ 
Bandwidth $\Delta\nu_{\rm total}$ (MHz) & 32 \\ 
Receiver temperature $T_{\rm rec}$ (K) & 30 \\ 
Telescope latitude $\phi$ & $+49.3^\circ$ (DRAO) \\
Phase-centre declination $\delta$ & $+49.3^\circ$ (zenith drift) \\
Beam model & Gaussian \\ 
Foreground model & Horizon wedge \\
& +buffer \\
Horizon buffer ($B$) & $0.1\,h\,{\rm Mpc}^{-1}$ \\
\hline\hline 
\end{tabular}
\caption{Instrumental and foreground parameters used to model the CHORD-like thermal-noise power spectrum with \texttt{py21cmSense}. The forecast is evaluated separately at each redshift using $\nu=\nu_{21}/(1+z)$.}
\label{tab:chord_params}
\end{table}

\section{Mesh-resolution convergence of inferred bias parameters}
\label{app:convergence}

The field-level bias-inference pipeline requires depositing both the \HI\ overdensity and the shifted operators onto a Cartesian mesh of side $N_{\rm grid}$. Because the transfer functions $\widehat{\beta}_a(k)$ are estimated as ratios of cross spectra (Eq.~\ref{eq:beta_estimator}), they can in principle carry a residual dependence on the mesh resolution: a coarser grid suppresses small-scale modes and modifies the high-$k$ shape of the transfer functions through the CIC window function and aliasing. The bias parameters themselves are extracted from the low-$k$ plateaus of the transfer functions (Eqs.~\ref{eq:beta_fit_lin}--\ref{eq:bias_from_beta}), so any mesh dependence should be confined to large $k$ and should not propagate into the inferred EFT parameters $(b_1,b_2,b_{\mathcal G_2},b_3)$. We verify this explicitly here.

\begin{figure*}
    \centering
    \includegraphics[width=\linewidth]{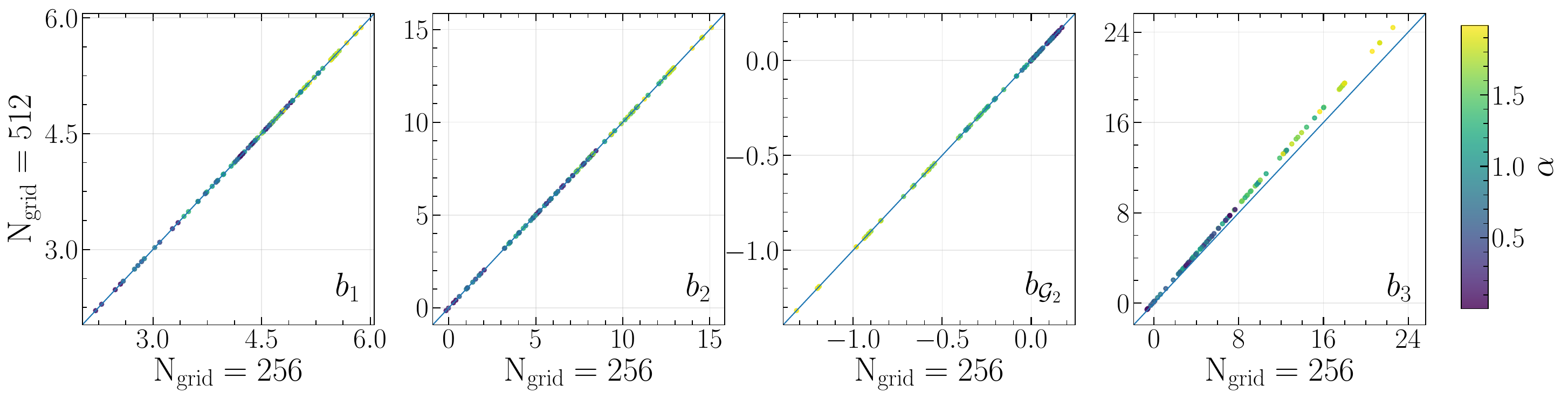}
    \caption{Mesh-resolution convergence of inferred bias parameters at $z=3$.
    Each panel compares one bias parameter measured with $N_{\rm grid}=256$ ($x$-axis) to the same measurement with $N_{\rm grid}=512$ ($y$-axis) for matched Hidden Valley HOD realisations, color-coded by $\alpha$.
    The diagonal blue line indicates perfect agreement.
    The first three parameters lie tightly on the diagonal across the full dynamic range of the HOD scan, while $b_3$ exhibits a mild discrepancy at higher values, confirming that the inferred large-scale bias coefficients are largely insensitive to the mesh resolution used in the field-level pipeline.}
    \label{fig:convergence}
\end{figure*}

Figure~\ref{fig:convergence} shows a direct comparison of the four inferred bias parameters obtained with $N_{\rm grid}=256$ and $N_{\rm grid}=512$ at $z=3.0$, for the sub-set of Hidden Valley HOD realisations used in Figs.~\ref{fig:hv_tng_comp_10_13}--\ref{fig:hv_tng_comp_lt12}. Each point corresponds to one HOD realisation, color-coded by $\alpha$.

For $b_1$, $b_2$, and $b_{\mathcal{G}_2}$ and the full range of HOD realisations, the $N_{\rm grid}=256$ and $N_{\rm grid}=512$ results lie tightly on the diagonal with no visible systematic offset or $\alpha$-dependent trend, while we see a mild discrepancy for $b_3$. While the two grid choices do produce slightly different transfer-function shapes at large $k$---where the CIC window function suppression and aliasing contributions differ between resolutions---these high-$k$ differences are absorbed by the smooth polynomial fitting forms in Eqs.~\eqref{eq:beta_fit_lin}--\eqref{eq:beta_fit_nonlin} and do not shift the extrapolated $k\to 0$ limits. The bias parameters are therefore a robust output of the pipeline, stable against the mesh resolution within the range tested.

For computational efficiency, all results in the main text use $N_{\rm grid}=256$, which gives sufficient sampling of the quasi-linear scales from which the large-scale bias limits are extracted, while keeping the memory and runtime requirements of the full HOD ensemble tractable.

\section{Comparison with \texttt{TNG300-Dark}}
\label{app:tngdark}

\begin{figure*}
\centering
\includegraphics[width=\linewidth]{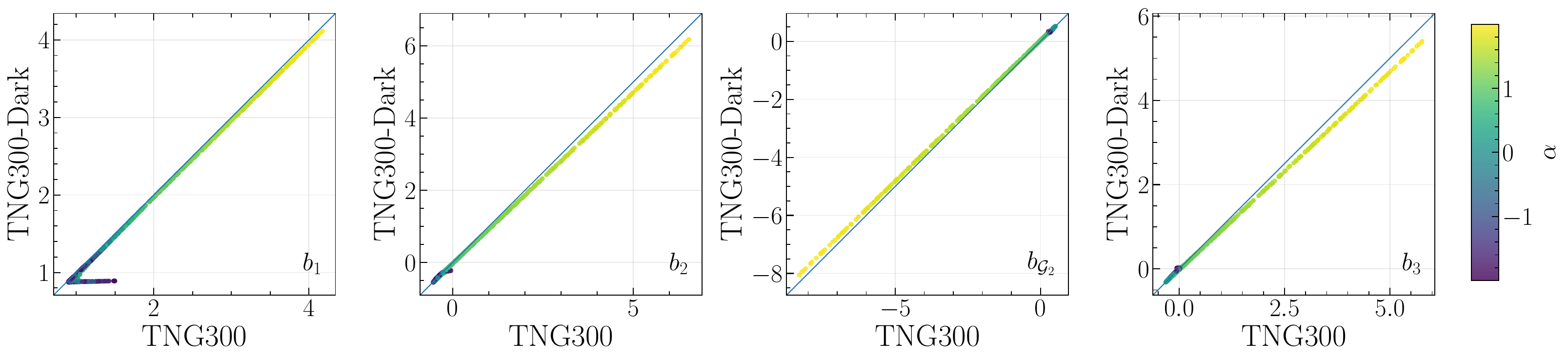}
\caption{Comparison of inferred EFT bias parameters at $z=1$ from the full-physics \texttt{TNG300} simulation and its dark-matter-only counterpart, \texttt{TNG300-Dark}, for matched HOD realizations. Each panel shows one bias parameter, with the diagonal line indicating equality between the two simulations. Points are color-coded by $\beta$. The close agreement indicates that baryonic effects in \texttt{TNG300} are not the dominant source of the offsets between Hidden Valley and \texttt{TNG300}.}
\label{fig:tng_dark_comp}
\end{figure*}

To test whether the offsets between Hidden Valley and \texttt{TNG300} are primarily driven by baryonic physics, we repeat the bias-inference procedure using the dark-matter-only counterpart of the simulation, \texttt{TNG300-Dark}. Figure~\ref{fig:tng_dark_comp} compares the inferred EFT bias parameters from \texttt{TNG300} and \texttt{TNG300-Dark} for matched HOD realizations, with points color-coded by the cutoff-sharpness parameter $\beta$.

The two TNG runs agree closely for all four bias parameters. Small deviations are visible, but the points largely follow the one-to-one relation. Importantly, the high-$\beta$ realizations do not show a strong separation between \texttt{TNG300} and \texttt{TNG300-Dark}. This suggests that the HV--TNG offsets seen in Sec.~\ref{sec:hv_tng_masssplit} are unlikely to be caused mainly by baryonic feedback in \texttt{TNG300}. They are more plausibly associated with other differences between the simulation suites, such as volume, halo abundance, mass calibration, resolution, or gravity-solver effects.

\bibliography{ref}

\end{document}